\documentclass[twocolumn,tighten]{aastex61}
\usepackage{color}
\usepackage{natbib}
\usepackage{amsmath}
\usepackage{comment}

\newcommand{\mdot}{\dot{M}}
\newcommand{\lnu}{L_{\nu}}
\newcommand{\ye}{Y_{\rm e}}
\newcommand{\bs}{\boldsymbol}
\newcommand{\pa}{\partial}
\newcommand{\pdrv}[2]{\frac{\partial #1}{\partial #2}}

\newcommand{\rd}{{\rm d}}

\newcommand{\beq}{\begin{equation}}
\newcommand{\eeq}{\end{equation}}
\newcommand{\beqa}{\begin{eqnarray}}
\newcommand{\eeqa}{\end{eqnarray}}
\newcommand{\ipro}{{\bs \cdot}}

\shorttitle{A parametric study of the acoustic mechanism}
\shortauthors{Harada et al.}

\begin{document}

\title{A Parametric Study of the Acoustic Mechanism for Core-Collapse Supernovae}

\author{A. Harada}
\affil{Physics Department, University of Tokyo, 7-3-1 Hongo, Bunkyo, Tokyo 113-0033, Japan}
\email{harada@utap.phys.s.u-tokyo.ac.jp}

\author{H. Nagakura}
\affil{TAPIR, Walter Burke Institue for Theoretical Physics, Mailcode 350-17, California Institute of Technology, Pasadena, CA 91125, USA}

\author{W. Iwakami and S. Yamada}
\affil{Advanced Research Institute for Science and Engineering, Waseda University, 3-4-1 Okubo, Shinjuku, Tokyo 169-8555, Japan}

\begin{abstract}
We investigate the criterion for the acoustic mechanism to work successfully in core-collapse supernovae. 
The acoustic mechanism is an alternative to the neutrino-heating mechanism. It was proposed by Burrows et al., who claimed that acoustic waves emitted by $g$-mode oscillations in proto-neutron stars (PNS) energize a stalled shock wave and eventually induce an explosion. Previous works mainly studied to which extent the $g$-modes are excited in the PNS. In this paper, on the other hand, we investigate how strong the acoustic wave needs to be if it were to revive a stalled shock wave. By adding the acoustic power as a new axis, we draw a critical surface, an extension of the critical curve commonly employed in the context of neutrino heating. We perform both 1D and 2D parametrized simulations, in which we inject acoustic waves from the inner boundary.
In order to quantify the power of acoustic waves, 
we use the extended Myers theory to take neutrino reactions into proper account. 
We find for the 1D simulations that rather large acoustic powers are required to relaunch the shock wave, since the additional heating provided by the secondary shocks developed from acoustic waves is
partially canceled by the neutrino cooling that is also enhanced. 
In 2D, the required acoustic powers are consistent with those of Burrows et al. Our results seem to imply, however, that it is the sum of neutrino heating and acoustic powers that matters for shock revival.
\end{abstract}


\section{Introduction}
\label{sec:intro}
Core-collapse supernovae (CCSNe) are 
deaths of massive stars. They are initiated by the gravitational collapse of an iron core at the end of the evolution 
and conclude with the explosion of outer envelopes with an energy of $\sim 10^{51}\,{\rm erg}$, which is accompanied with the formation of compact objects such as neutron stars and black holes. 
The explosion is assumed to be produced by the passage of the shock wave generated by the core bounce, which occurs when the central density exceeds the nuclear saturation density. The long-standing problem is that the shock wave stalls inside the core, 
consuming its energy to dissociate 
nuclei, and it cannot be clearly determined how 
this once-stalled shock is revived and eventually produces an explosion.

The current leading hypothesis 
is the neutrino-heating mechanism, in which the 
energy of $\sim 10^{53}\,{\rm erg}$ that is stored in the central portion of the core, or the proto-neutron star (PNS), 
is emitted in the form of 
neutrinos, and a fraction of the neutrinos is reabsorbed to 
heat and revive the shock.

According to recent studies 
\citep[see][for a review]{2016ARNPS..66..341J}, multidimensional fluid instabilities such as neutrino-driven convection and standing accretion shock instability (SASI) are crucially important for the scenario: 
these instabilities induce turbulence and push the shock outward with thier turbulent pressure 
\citep{2013ApJ...771...52M, 2015ApJ...799....5C}; lateral motions in the turbulence 
increase the dwell time of fluid elements in the so-called gain region, where heating dominates cooling \citep{2012ApJ...749...98T}; in multidimensions, 
both upward motions of hot and buoyant matter and down-flows of cold material are realized simultaneously, which enhances the efficiency of heating.

All these effects combined have recently produced successful shock revivals in many 2D and some 3D simulations, which failed earlier in 1D under spherical symmetry \citep{2000ApJ...539L..33R, 2001PhRvD..63j3004L, 2003ApJ...592..434T, 2005ApJ...629..922S}. 
Some problems remain in these seemingly successful models, however. For one, results from different groups appear to be at odds with one another \citep{2012ApJ...749...98T, 2013ApJ...767L...6B, 2015ApJ...800...10D, 2014ApJ...786...83T, 2014PhRvD..90d5032T, 2015MNRAS.453..287M}. 
For another, the explsion energy obtained in these simulations is commonly lower than the canonical value of $\sim 10^{51}\,{\rm erg}$ 
\citep[see, however,][]{2016ApJ...818..123B}. These problems may imply that some important physical process(es) is (are) still missing in the neutrino-heating mechanism.

The acoustic mechanism was 
proposed by \citet{2006ApJ...640..878B, 2007PhR...442...23B, 2007ApJ...655..416B} as an alternative to the neutrino-heating mechanim. 
In the context explained above, the ``missing physics" is the oscillation of 
PNS. 
In this mechanism, 
turbulent accretion flows caused by SASI and/or convection beat 
the PNS anisotropically and repeatedly. These impulsive forces excite 
$g$-mode oscillations of PNS, which then 
emit acoustic waves into the accreting matter. The waves are steepened to form 
secondary shock waves as they propagate outward, and they deposit 
energy to shocked matter. 
The PNS in this way transduces the gravitational energy of the accretion flow into the energy of acoustic waves. This conversion of energy was claimed to last 
until an 
explosion is instigated, since the matter accretion, the energy source for acoustic waves, also continues until the explosion occurs. 
Another 
merit of the acoustic waves that the Burrows' group claimed is that the waves do not go beyond the shock wave and hence deposit all their energies inside it. 

One of the key issues in the acoustic mechanism is to what extent the oscillations of PNS are excited. 
In their papers, 
\citet{2006ApJ...640..878B, 2007PhR...442...23B, 2007ApJ...655..416B} found numerically that the energies of the excited 
$g$-modes are 
$10^{50}\text{--}10^{51}\,{\rm erg}$ with the 
$\ell = 1$ mode being the most pronounced with a period of $\sim 3\,{\rm ms}$, where $\ell$ is the degree 
of the spherical harmonics. They also observed that these oscillations of PNS emit acoustic waves with powers 
of $\sim 10^{51}\,{\rm erg\,s^{-1}}$. 
Unfortunately, these results have not been reproduced in the simulations by other groups \citep{2009ApJ...694..664M}. Although in simulations of \citet{2009ApJ...694..664M} the small central region is treated in 1D, they claimed and checked that the spherically symmetrized region was so small that their simulations could reproduce the core $g$-mode oscillation if it really occurred. It is true that such differences in numerical methods between Burrows' group and others may not be ignored, but it is unclear why other groups fail to reproduce strong $g$-mode oscillations. 
Long-term computations required to obtain acoustic waves also hamper systematic studies by realistic simulations.

\citet{2007ApJ...665.1268Y} took a different, more phenomenological approach based on 
linear perturbation theory. They calculated forced oscillations of PNS by 
pressure fluctuations that are possibly imparted to 
the PNS surface by SASI. Employing the results of the numerical simulation of SASI by \citet{2006ApJ...641.1018O}, they estimated that the energies of the excited $g$-modes are 
$\la 10^{50}\,{\rm erg}$ and argued that these forced oscillations would inject energy at a 
rate of $\sim 10^{51}\,{\rm erg\,s^{-1}}$ as acoustic waves, which is comparable to what 
\citet{2006ApJ...640..878B} obtained. 

Then came a serious challenge from 
\citet{2008MNRAS.387L..64W}, who analyzed the nonlinear 
three-mode couplings among $g$-modes, which 
transfer energy to two 
``daughter" modes with lower frequencies. 
The wave energies are assumed to be eventually 
dissipated and emitted as neutrinos. 
According to their calculations, the excitation of the most pronounced mode with 
$\ell = 1$ is saturated at 
$\sim 10^{47-48}\,{\rm erg}$ for the steady energy feed at 
$\dot{E} = 10^{50-51}\,{\rm erg\,s^{-1}}$ from the matter accretion, which, if we assume 
the acoustic damping rate of 
$10\,{\rm Hz}$, gives the acoustic power of 
$\sim 10^{48}\text{--}10^{49}\,{\rm erg\,s^{-1}}$. These values are 
$\sim 10\text{--} 100$ times lower than the values given in \cite{2006ApJ...640..878B}. We note that the numerical simulations by the latter authors have most likely failed to capture these mode couplings, 
since one of the daughter 
modes of relevance for resonant 
couplings has such short wavelengths that the numerical resolutions were not sufficient. It should be mentioned that \citet{2007ApJ...665.1268Y} did not take the mode coupling into account, either. This is certainly a serious issue, but we recall that 
\citet{2008MNRAS.387L..64W} also made some assumptions in their analysis. For instance, although they assumed a steady 
energy injection from turbulent accretion flows to $g$-modes, this may not be a good approximation, since 
the forces excerted on 
the PNS suraface are more like 
a collection of impulsive 
hits, as envisaged by 
\citet{2007ApJ...665.1268Y}. 
It also remains to be confirmed if the mode couplings neglected in their paper are indeed minor. Investigations by other independent groups are certainly desirable. 

It is true that the best way is in principle either to improve the numerical resolution sufficiently or to conduct a fully nonlinear analysis of the mode couplings, but this is almost impossible for the moment. We hence take yet another way in this paper. 
Reversing the argument, we ask here what acoustic power is needed to revive the stalled shock wave. 
This is in accord with the spirit of 
the critical curve theory for the neutrino-heating mechanism \citep{1993ApJ...416L..75B}, in which the 
critical neutrino luminosity for shock revival is considered 
as a function of the mass accretion rate. 
In this paper we 
add a new dimension in this theory and discuss 
the critical ``surface." Although the rotational velocity was introduced 
by \citet{2014ApJ...793....5I} to consider such a 
critical surface, the new dimension introduced in this paper is 
the intensity of acoustic waves. Then the critical surface will allow us to assess how intense acoustic waves need to be to obtain 
shock revival, which will in turn help us judge how promising 
the acoustic mechanism is 
in realistic settings. 
We note that in this framework of the critical surface we are in fact treating the acoustic and neutrino powers on an equal footing, which may be in contrast to the original claim by Burrows et al. that the acoustic waves are the dominant agent for shock revival. In this sence, our model may be referred to as a ``hybrid" model.

In order to obtain the critical surface, 
we perform both 
1D and 2D simulations under spherical and axial symmetries, respectively. 
Since the $g$-modes are non-spherically symmetric and hence the acoustic mechanism 
works only in multidimensions, the 
1D simulations may not be realistic. Spherical acoustic waves are still conceivable, however, and they will be easier to analyze than non-spherical conterparts and hence are useful to capture the essential features in the propagation and energy deposit of acoustic waves. 

This paper is organized 
as follows. 
In section \ref{sec:methods} we describe the numerical methods. The results of simulations are shown in the subsequent sections. In section \ref{sec:1dresult} we first summarize 1D models. Then 
2D results are presented 
in section \ref{sec:2dresult}. In section \ref{sec:concl} we give 
summary and some discussions.

\section{Methods}
\label{sec:methods}
In our models we prepare spherically symmetric 
steady accretion flows with a constant mass accretion rate $\mdot$ and 
neutrino luminosity $\lnu$ and inject 
acoustic waves continuously from the inner boundary of the computation domain, which is located close to the 
PNS surface. 
For 
various combinations of $\mdot$ and $\lnu$, we search the critical amplitudes of acoustic wave, which are the minimum amplitudes required for shock revival. 
We regard it as a successful 
shock revival 
if the shock reaches the radius of 
$500\,{\rm km}$ within $500\,{\rm ms}$ from the onset 
of the acoustic wave injection. We then draw 
the critical surface in the space spanned by $\mdot$, $\lnu$, and the amplitude of the acoustic wave.

Basic equations in our calculations are inviscid hydrodynamics equations with neutrino emissions and absorptions as an energy sink/source and an equation for the electron fraction: 
\beq
\pdrv{\rho}{t} + {\bs \nabla}\ipro (\rho {\bs v}) = 0, \label{eq:continuity}
\eeq
\beq
\pdrv{\rho {\bs v}}{t} + {\bs \nabla} \ipro (\rho {\bs v \bs v} + P {\bs I}) = -\rho {\bs \nabla}\Phi, \label{eq:momentum}
\eeq
\beqa
\pdrv{\rho ( e + \onehalf{\bs v}^2 )}{t} &+& {\bs \nabla}\ipro \left\{\rho {\bs v}\left(e + \frac{1}{2}{\bs v}^2 + \frac{P}{\rho} \right)\right\} \nonumber \\ &=& - \rho {\bs v}\ipro {\bs \nabla}\Phi + Q, \label{eq:energy}
\eeqa
\beq
\pdrv{\rho \ye}{t} + {\bs \nabla}\ipro(\rho {\bs v}\ye) = \rho \Gamma, \label{eq:electronfraction}
\eeq
\beq
\Phi = - \frac{GM_{\rm PNS}}{r}, \label{eq:pointmass}
\eeq
where $\rho$, $\bs v$, $P$, $e$, $Q$, $\ye$, $\Gamma$, $\Phi$, $G$, $M_{\rm PNS}$, and $r$ are the density, velocity, pressure, specific internal energy, 
net heating rate via neutrino emissions and absorptions, electron fraction, rate of change in $\ye$ by the neutrino reactions, 
gravitational potential by 
PNS, gravitational constant, mass of 
PNS, and distance from the center, respectively; 
$\bs I$ and 
$\bs v \bs v$ are the unit and dyadic tensors, respectively; the self-gravity of the accreting matter is ignored, and only the gravitational attraction by PNS is considered. In the following 
calculations, the PNS mass is fixed to $M_{\rm PNS} = 1.4\,M_\odot$. We employ 
the so-called STOS equation of state (EOS) 
\citep{1998NuPhA.637..435S}, which is based on the relativistic mean field theory and the Thomas-Fermi approximation. 
The light-bulb method in \cite{2006ApJ...641.1018O}\footnote{We drop 
$\pi$ from the denominator of 
equation (18) in their paper.} is adopted 
to calculate $Q$ and $\Gamma$ in equations (\ref{eq:energy}) and (\ref{eq:electronfraction}). We consider only 
the 
absorption and emission of neutrinos by free nucleons, $\nu_{\rm e} + {\rm n} \leftrightarrow {\rm e}^- + {\rm p}$ and $\bar\nu_{\rm e} + {\rm p} \leftrightarrow {\rm e}^+ + {\rm n}$. 
The neutrino temperatures 
are set to 
$T_{\nu_{\rm e}} = 4\,{\rm MeV}$ for $\nu_{\rm e}$ and $T_{\bar \nu_{\rm e}} = 5\,{\rm MeV}$ for $\bar \nu_{\rm e}$.

We run 1D and 2D simulations on the spherical coordinates under spherical and axial symmetries, respectively. 
The inner boundary of the computational domain $r_0$ is fixed to the neutrinosphere $r_\nu$, which is defined in this paper to be the 
radius at 
the density of 
$10^{11}\,{\rm g\,cm^{-3}}$ in the initial condition, which is explained below. 
The radial mesh width $\Delta r_i$ at the $i$th grid point is set to $1\%$ of the radius $r_i$: $\Delta r_i = 0.01 r_i$. 
The number of radial grid points is  
$256$, but if the outermost radius $r_{256}$ is smaller than $500\,{\rm km}$, we increase the number to $320$ so that $r_{320}$ exceeds $500\,{\rm km}$. 
In 2D models, the entire meridian section is covered with the same radial grid points and $128$ $\theta$-grid points, the latter of which are deployed according to 
the Gaussian quadrature points and weights 
as in \cite{2012ApJS..199...17S}.

The hydrodynamical code employed in this paper is the same as that in \cite{2014ApJS..214...16N}, except that only the point-mass gravity instead of the 
self-gravity is considered: 
the Harten-Lax-van Leer scheme \citep{1983SIAMR...25...35H} with the piecewise parabolic interpolation \citep{Colella1984174} is used to evaluate the numerical flux; the 
time evolution 
is handled 
by the explicit, total-variation diminishing, third-order Runge-Kutta method.

The initial conditions are time-independent 
solutions of equations 
(\ref{eq:continuity})--(\ref{eq:pointmass}) for given combinations of 
constant mass accretion rate and 
neutrino luminosity. The 
numerical method to obtain these solutions 
is essentially 
the same 
as that in \citet{2006ApJ...650..291Y}, except for two aspects: one 
is again the 
use of the point-mass gravity instead of the self-gravity, and the 
other is the definition of the 
neutrinosphere $r_\nu$, which in this paper is defined to be the radius at which the density is $10^{11}\,{\rm g\,cm^{-3}}$, wheres in \citet{2006ApJ...650..291Y} it was 
the radius, where the optical depth is $2/3$. Regardless, 
the neutrino luminosity 
is assumed to satisfy 
the following relation at the neutrinosphere: 
$L_\nu = \frac{7}{16} \sigma T_{\nu_{\rm e}}^4 4\pi r_\nu^2$. In the upstream of the standing shock wave, on the other hand, we assume that matter is accreting spherically symmetric with the 
entropy per nucleon $s$, 
electron fraction $Y_{\rm e}$, and radial velocity $v_{r}$ being $s = 3$ in units of the Boltzmann constant $k_{\rm B}$, $Y_{\rm e} = 0.5$, and $v_r = \sqrt{2GM/r}$, respectively.

We inject acoustic waves from the inner boundary. We impose time-dependent boundary conditions there to generate outgoing sound waves. 
We assume a 
sinusoidal oscillation in 
the density, $\rho = \rho_0 + \rho_1 \sin(\omega t - kr)$, where $\rho_0$ is the density of the steady state 
at the inner boundary, 
whereas $\rho_1$, $\omega$, and $k$ are 
the amplitude, 
frequency, and 
wavenumber of the oscillation, respectively. 
We assume $\rho_1 \propto \mathcal{P}_0(\mu)$, where $\mathcal{P}_\ell$ is the Legendre polynomial of $\ell$-th order and $\mu = \cos \theta$, in 1D simulations. Although $g$-mode oscillations, the main source of the acoustic waves, are intrinsically non-spherical, we consider 1D spherically symmetric acoustic waves with $\ell = 0$ as well in this paper, since they elucidate the essential feature of the acoustic energy transport. In 2D simulations, on the other hand, we set $\rho_1 \propto \mathcal{P}_1(\mu)$, although we can consider any non-zero $\ell$ in principle, since $\ell = 1$ modes were the most prominent in \citet{2006ApJ...640..878B}. 
Throughout this paper, the normalized dimensionless 
amplitude of acoustic wave $\delta$ is defined as 
$\rho_1 = \rho_0 \mathcal{P}_\ell \delta$ for $\ell=0$ (1D) and $\ell=1$ (2D). The entropy per 
nucleon $s$ and electron fraction $\ye$ at the inner boundary are fixed to the steady-state values, reflecting the adiabatic character of acoustic waves. Other thermodynamic quantities such as temperature and pressure are determined by the EOS. 
The inner boundary condition for velocity is determined so that it should be consistent with the outgoing sound waves: $v_r=v_0 + a \mathcal{P}_\ell \delta \sin(\omega t - kr)$, 
where $v_0$, $a = \sqrt{(\partial P/\partial \rho)_{s,Y_{\rm e}}}$, and $k=\omega/(v_0 + a)$ are the velocity, sound speed, and wave number at the inner boundary in the 
steady state, respectively. 
The frequancy is set to $\omega = 2\pi/(3\,{\rm ms})$, i.e., the oscillation period is 
$3\,{\rm ms}$, the value of the dominant $g$-mode oscillation 
in \citet{2006ApJ...640..878B}. Incidentally, the values of all quantities at the outer boundary are fixed to the values in the steady state.

For each combination of 
$\mdot$, $\lnu$, and 
$\delta$, we run a simulation 
for $500\,{\rm ms}$. If the mean shock radius exceeds $500\,{\rm km}$ within this period, we interpret it as shock revival and consider that this 
model produces a successful explosion. This value of the radius is the same as the value adopted in \citet{2014ApJ...793....5I} and slightly more conservative 
compared with the value of $400\,{\rm km}$ employed in 
\citet{2010ApJ...720..694N} and \cite{2012ApJ...755..138H}. 
The shock radius is defined in this paper as 
the radius at which 
the 
entropy per nucleon is $s = 6\,k_{\rm B}$. 
This is roughly twice 
the entropy in 
the unshocked accretion flow, as explained above. We vary 
$\delta$ at regular 
intervals of $0.005$ (1D) or $0.01$ (2D), 
and search for the threshold of $\delta$ for shock revival for each combination 
of $\mdot$ and $\lnu$. Connecting the points obtained this way, we draw 
the critical surface, a 2D analogue of the critical curve.

Before considering the results, 
we mention the 
dependence on 
the numerical resolution and the initial phase of the acoustic wave. 
In order to assess these effects, we conducted 
additional simulations. We first reduced the 
radial grid width by half in the 1D simulations with the 
mass accretion rate 
$\dot{M}=0.6\,M_\odot\,{\rm s^{-1}}$ and neutrino luminosity 
$\lnu=4.0\times 10^{52}\,{\rm erg\,s^{-1}}$ 
and found that 
the critical surface is shifted 
only by $0.01\text{--}0.015$ in the positive direction of the $\delta$ axis. 
As for the choice of the initial phase of the oscillation, we ran four additional computations for the same 1D model with the original resolution, but with the phase being changed by $\pi/2$, and we 
confirmed that 
the critical surface is shifted only by $0.005$ in $\delta$. 
We thus concluded that 
the critical surface in 
1D is determined fairly unambiguously. 
We also checked 
the numerical resolution 
in 2D simulations and confirmed that it is sufficient. 
In fact, doubling the number of grid points in the radial (angular) direction in the 
simulations with $\dot{M} = 1.0 \,M_\odot\,{\rm s^{-1}}$ and $\lnu = 4.5\times 10^{52}\,{\rm erg\,s^{-1}}$ lowered 
the critical surface 
only by $\sim0.02$ ($\sim 0.01$) in $\delta$.
Moreover, the change in 
initial phase by $\pi/2$ in the same model 
shifts the critical surface only by $\sim 0.01$ in $\delta$. This shows 
that the critical surface is probably also well determined in 2D.

\section{Results in 1D}
\label{sec:1dresult}
\subsection{Critical Surface}
\label{sec:1dcritical}
\begin{figure}[tbp]
\centering
\includegraphics[width=\hsize]{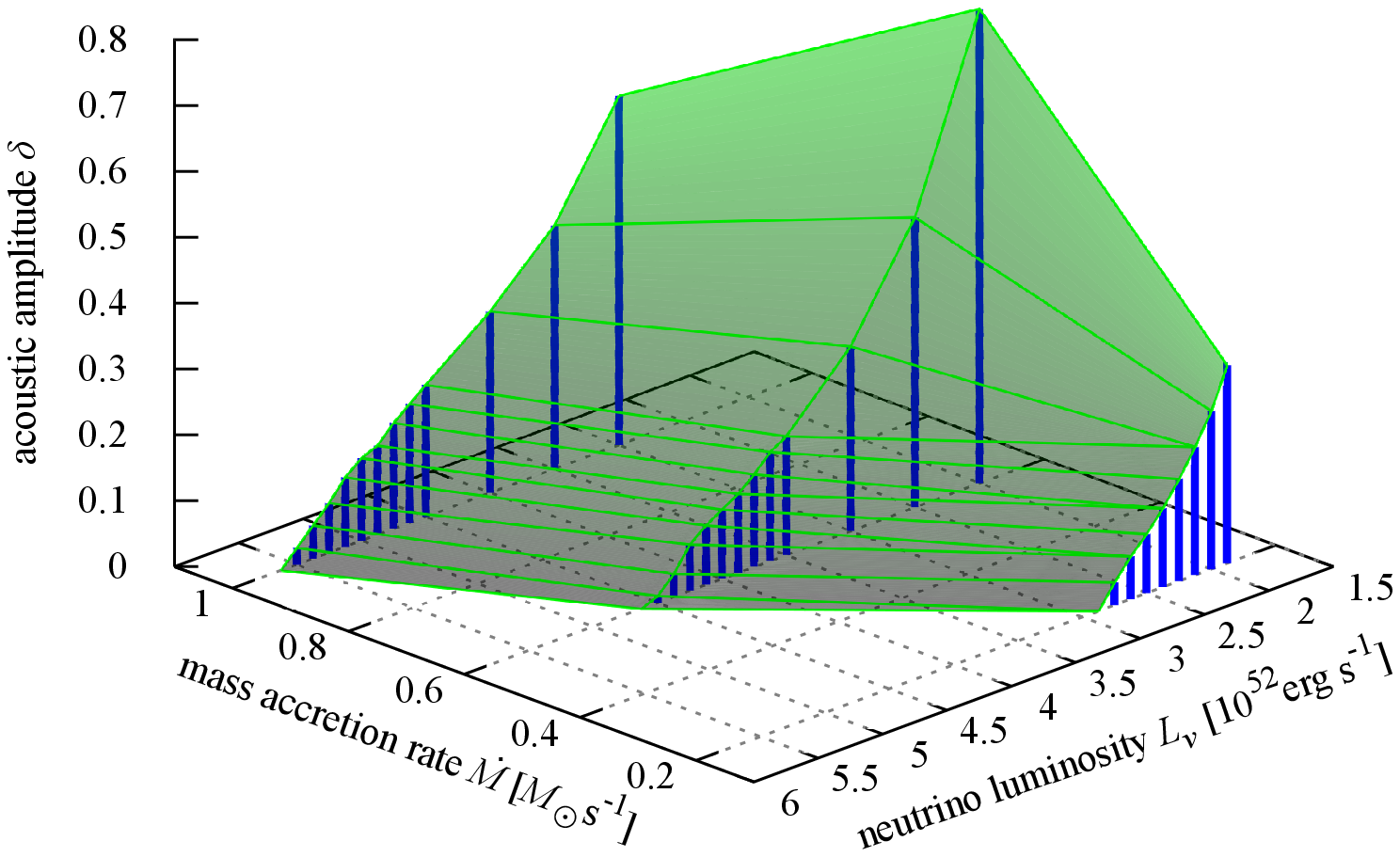}\\
\includegraphics[width=\hsize]{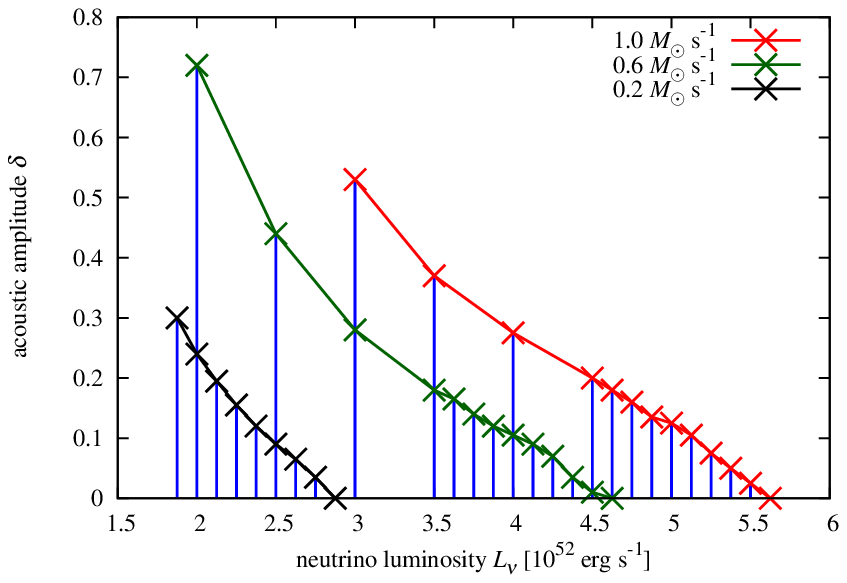}
\caption{\label{fig:1dcrit} (Upper panel) Critical surface 
for 1D 
models in the $\dot{M}\text{--}\lnu\text{--}\delta$ space. Models with parameters on 
or above the surface 
produce explosions. Blue dropping vertical lines from the surface indicate $\mdot$ and $\lnu$ 
adopted in the models. (Lower panel) Lines on the critical surface for constant mass accretion rates projected onto the $\lnu$--$\delta$ plain. 
Blue lines are the same as those in the upper panel.}
\end{figure}

The critical surface we obtained from the 1D simulations is shown in the upper panel of Figure \ref{fig:1dcrit}. 
By definition, models 
with the parameters on 
or above 
this surface 
result in shock revival, whereas those beneath it fail.  
In the lower panel, three lines on the critical surface, each of which has an 
identical mass accretion rate, 
are projected onto the $\delta \text{--} L_\nu$ plain. 
Before discussing these results in detail, we first consider a representative model and examine its evolution.

\begin{figure}[tbp]
\includegraphics[width=\hsize]{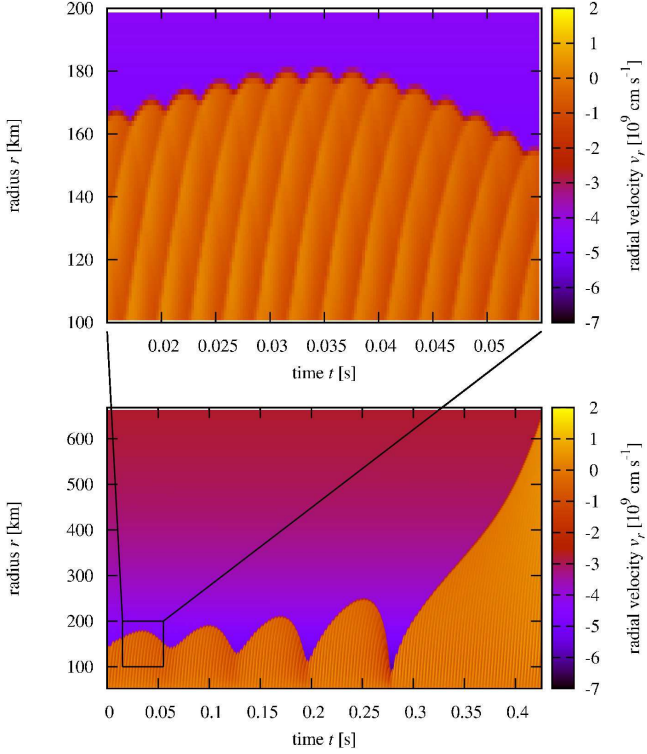}
\caption{\label{fig:m06l40a105} Time evolution of the radial 
velocities for the 
model just on the critical surface with $\mdot = 0.6\,M_\odot\,{\rm s^{-1}}$, $\lnu = 4.0 \times 10^{52}\,{\rm erg\,s^{-1}}$, and $\delta = 0.105$. 
The boundaries between different colors 
represent shock waves. The upper panel is a zoom-in 
of the designated area in the lower panel. 
}
\end{figure}

In Figure \ref{fig:m06l40a105} we show the 
temporal evolution of the radial velocity in the model with $\mdot = 0.6\,M_\odot\,{\rm s^{-1}}$, $\lnu=4.0 \times 10^{52}\,{\rm erg\,s^{-1}}$, and $\delta = 0.105$ 
as an example. 
This model is located on the critical surface. Discontinuous changes in color in the figure correspond to shock waves. We can clealy see radial oscillations of the primary-shock wave with a period of $\sim 70\,{\rm ms}$ in the lower panel. This is an oscillatory 
instability found by 
\citet{2012ApJ...749..142F}, since the 
timescale of the advection from the shock to the point of maximum cooling is $\sim 24\,{\rm ms}$, 
indeed 
between $1/4$--$1/2$ 
of the period as expected in his paper. 
The upper panel, on the other hand, is a zoom-in to the designated area in the lower panel. It is evident in this panel that many shock waves are propagating outward on top of the 
oscillatory mode and periodically hit the primary-shock wave. 
These secondary shock waves 
are generated as a consequence of the steepening of acoustic waves injected from the inner boundary. This is understood from the fact that the interval between the consecutive shock waves is exactly equal to the period of the acoustic wave. Although these 
shocks are rather weak (the typical Mach number is $\sim 1.3$) 
in this model, each collision of a secondary shock with the primary shock causes a tremor in the latter, which is clearly visible in the upper panel. 

\begin{figure}[tbp]
\centering
\includegraphics[width=\hsize]{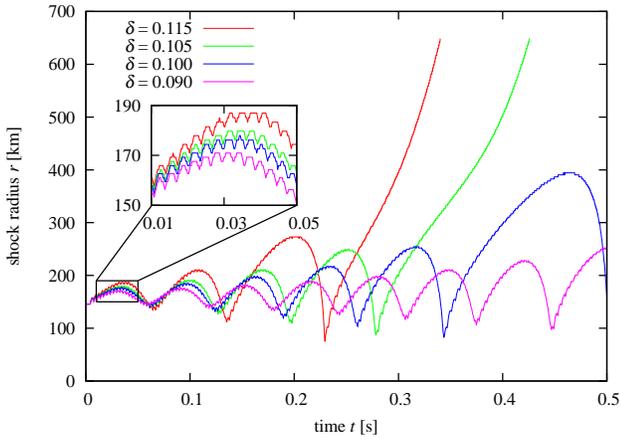}
\caption{\label{fig:shocks} Time variations of the primary-shock radii 
for some of the models with $\mdot = 0.6\,M_\odot \,{\rm s^{-1}}$ and $\lnu = 4.0 \times 10^{52}\,{\rm erg\,s^{-1}}$. Both 
successful (red and green) and 
failed (blue and magenta) models are 
included. The model with $\delta = 0.105$ is on the critical surface. 
Shown in the 
inset is the zoom-in of 
particular period near the onset of the simulations.}
\end{figure}

The 
oscillatory motion of the primary-shock wave repeats itself with growing amplitudes. 
These overstable oscillations eventually lead to shock revival in this model, as 
demonstrated in the lower panel of Figure \ref{fig:m06l40a105}. Figure \ref{fig:shocks} shows 
other models with various acoustic amplitudes $\delta$. 
The inset is a zoom-in of the rectangular area in the main panel and displays tremblings caused by the secondary shock waves in these models. 
We can see from the figure that as the acoustic amplitude $\delta$ decreases, the time to shock revival increases, and eventually, no shock revival occurs, 
indicating that there is a critical amplitude for shock revival. This 
is common to other models with different combinations of $\dot{M}$ and $\lnu$. The collection of these critical amplitudes gives 
the critical surface shown in Figure \ref{fig:1dcrit}.

Although the acoustic amplitude is employed as the 
independent variable in the simulations, the acoustic power $\dot{E}_{\rm aco}$, which is the rate at which matter is heated by the acoustic wave and should be compared with the neutrino-heating rate, is 
more convenient for 
physical interpretations and comparisons with previous works \citep{2006ApJ...640..878B, 2007ApJ...665.1268Y, 2008MNRAS.387L..64W}. 
Since acoustic waves do not cross 
the shock and hence cannot escape from the postshock region, 
the acoustic power $\dot{E}_{\rm aco}$ should be equal to the acoustic luminosity $L_{\rm aco}$ at the inner boundary, which is the surface integral of the acoustic energy flux. 
It is well known that the energy density and 
flux of an acoustic wave 
are proportional to the amplitude squared as long as its 
amplitude is small enough to be 
in the linear regime. 
In our models, however, the amplitudes 
are 
not small in general and the acoustic waves may not be regarded as linear waves. 
Moreover, the fact that the acoustic waves are propagating on top of matter that is not at rest, but flows non-uniformly, complicates the evaluation of acoustic power even more. In order to 
handle these problems, we extend the Myers theory \citep{MYERS1986277, 1991JFM...226..383M}. 

Myers derived the corollary of the energy conservation 
for disturbances 
in homentropic flows, i.e., acoustic waves \citep{MYERS1986277}, 
as well as in flows 
with inhomogeneous 
entropies \citep{1991JFM...226..383M}. 
Here 
we extend the discussion of \citet{1991JFM...226..383M} in order to 
take the effects of 
neutrino reactions into account. As 
derived in appendix \ref{sec:myers}, the 
resulting equation is expressed as follows: 
\beq
\pdrv{E_{\rm dis}}{t} + {\bs \nabla}\ipro{\bs F}_{\rm dis} = - D_{\rm dis}, \label{eq:cor}
\eeq
where each term is given as
\begin{eqnarray}
E_{\rm dis} &=& \rho \left(H-H_0 - T_0 (s-s_0) - \frac{\mu_0}{m_{\rm u}} (\ye - {\ye}_0)\right) \nonumber \\
&-& \bs m_0 \ipro (\bs u - \bs u_0) - (P-P_0), \label{eq:eaco}
\eeqa
\beqa
\bs F_{\rm dis} &=& (\bs m - \bs m_0) \left(H-H_0 - T_0 (s-s_0) - \frac{\mu_0}{m_{\rm u}} (\ye - {\ye}_0)\right) \nonumber \\
&+& \bs m_0 \left((T-T_0)(s-s_0) + \frac{\mu-\mu_0}{m_{\rm u}}(\ye - {\ye}_0)\right), \label{eq:faco}
\eeqa
\beqa
D_{\rm dis} &=&- (s-s_0) \bs m_0 \ipro \bs \nabla (T-T_0) - (\ye - {\ye}_0) \bs m_0 \ipro \bs \nabla \frac{\mu - \mu_0}{m_{\rm u}} \nonumber \\
+ (\bs m &-& \bs m_0) \ipro \left(\bs \zeta - \bs \zeta_0 + (s-s_0) \bs \nabla T_0 + (\ye - {\ye}_0) \bs \nabla \frac{\mu_0}{m_{\rm u}}\right) \nonumber \\
- (T&-&T_0)\left(\frac{Q}{T} - \frac{Q_0}{T_0}\right) + \frac{\mu \mu_0}{m_{\rm u}} \left(\frac{T}{\mu} - \frac{T_0}{\mu_0} \right) \left( \frac{\rho \Gamma}{T} - \frac{\rho_0 \Gamma_0}{T_0} \right). \nonumber \\
\label{eq:daco}
\end{eqnarray}
In the above equations, $m_{\rm u}$, $H = e + P/\rho + \onehalf {\bs v}^2$, ${\bs m}=\rho {\bs v}$, $T$, and $s$ are the atomic mass unit, specific stagnation enthalpy (or the Bernoulli function),
mass flux, temperature, and specific entropy, respectively; $\mu$ is the chemical potential of electron-type neutrinos defined with the chemical potentials of electrons, protons, and neutrons, $\mu_{\rm e,p,n}$,  
as $\mu = \mu_{\rm e} + \mu_{\rm p}-\mu_{\rm n}$; $\bs \zeta$ is defined as 
$\bs \zeta = \bs \omega \bs \times \bs v$, where $\bs \omega = \bs \nabla \bs \times \bs v$ is the vorticity. The quantities with and without subscript $0$ stand for the unperturbed and perturbed 
variables, respectively. 
Although it is difficult to give an unambiguous interpretation to each term of Eq. (6), we regard $E_{\rm dis}$, $\bs F_{\rm dis}$, and $D_{\rm dis}$ as the density, flux, and dissipation, respectively, of the energy of acoustic waves with not necessarily small amplitudes. This may be justified by the facts that they obey the equation of a conservative form and that they are reduced to the well-known counterparts for linear waves if the amplitude is small and all neutrino contributions are turned off; they are hence natural extensions. See appendix \ref{sec:myers} for more discussions.

\begin{figure}[tbp]
\centering
\includegraphics[width=\hsize]{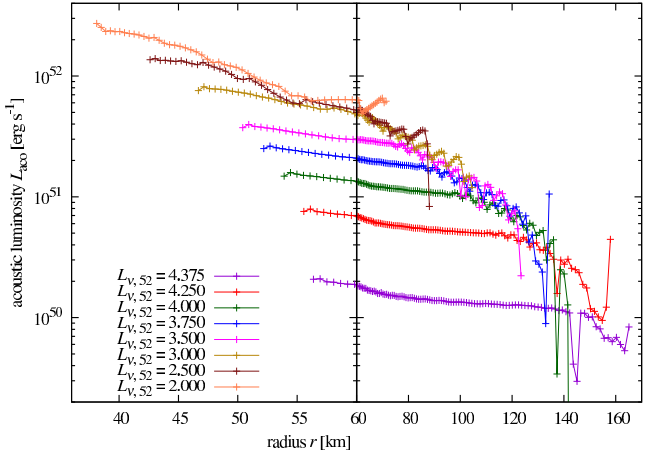}
\caption{\label{fig:fluxprofile} Radial profiles of the acoustic luminosities $L_{\rm aco}$ for selected models on the critical surface. The luminosities are averaged over 
the period of $3\,{\rm ms}$. The pluses 
present the on-grid values. The mass accretion rate $\mdot$ 
is $0.6\,M_\odot\,{\rm s^{-1}}$ for all models. Different lines represent 
the results for different neutrino luminosities $\lnu$, 
which have different critical acoustic amplitudes $\delta$. The 
normalized luminosity $L_{\nu,\,52}$ in the legend is defined 
as $L_{\nu}/(10^{52}\,{\rm erg\,s^{-1}})$.}
\end{figure}

One may think that the acoustic 
luminosity should be evaluated 
at the inner boundary, where the acoustic waves are generated artificially, but this may not be true, since the injected waves will be partially reflected back immediately after they enter the computational domain. This fact can be understood from Figure \ref{fig:fluxprofile}, which shows the acoustic luminosities defined at each radius as the surface integral of the radial component of Myers' flux, $L_{\rm aco}(r) = 4\pi r^2 F_{{\rm dis},r}(r)$, for models sitting on the critical surface. We note that the luminosity, $L_{\rm aco}(r)$, at the radius $r$ is obtained by taking its average over $3\,{\rm ms}$, the period of the acoustic wave, from the instant when the acoustic wave just reaches the radius. One can see small transients on the first two grid points in all cases. 
We hence decided to use the acoustic luminosity 
obtained at the 
third grid point from the inner boundary, where the initial adjustment appears to have been already over, to estimate the {\it truly} injected acoustic power $\dot{E}_{\rm aco}$. \citet{2012ApJ...749..142F} 
gave a similar argument that the third grid point is the innermost point that is not significantly affected by the inner boundary. 
We note that we can recognize a common trend in this figure that the acoustic luminosities 
decrease with 
radius. 
This is particularly significant for models with small $\lnu$'s. 
The reduction of the acoustic luminosity may be 
attributed to the dissipation term $D_{\rm dis}$ in equation (\ref{eq:cor}). 

It will be useful to give an order-of-magnitude estimate of the acoustic power here. 
Only the radial components 
are considered for vectors. 
The typical values of the temperature, specific entropy, electron fraction, chemical potential, specific stagnation enthalpy, 
and mass flux 
are $k_{\rm B} T_0 \sim {\rm MeV}$, $s_0 \sim 10 \,k_{\rm B}$ per nucleon, $Y_{\rm e0} \sim 0.1$, $\mu_0 \sim {\rm MeV}$, $H_0 \sim 10^{19}\,{\rm erg\,g^{-1}}$, and $m_{0r} \sim -10^{18\text{--}19}\,{\rm g\,cm^{-2}}$, respectively, at $r\simeq 50\,{\rm km}$. 
As for the disturbances, 
the amplitudes are typically 
$\sim 10\%$ of the unperturbed counterparts, except for the mass flux, for which $m_r - m_{0r}$ is $\sim 1\text{--}10$ times larger than $m_{0r}$ and is positive. 
Then,
\beq
(m_r - m_{0r})(H-H_0) \sim 10^{36\text{--}38}\,{\rm erg\,cm^{-2}\,s^{-1}}.
\eeq
Similarly,
\beqa
(m_r-m_{0r}) T_0 (s-s_0) &\sim& 10^{36\text{--}38}\,{\rm erg\,cm^{-2}\,s^{-1}}, \\
m_{0r} (T-T_0)(s-s_0) &\sim& 10^{35\text{--}36}\,{\rm erg\,cm^{-2}\,s^{-1}},
\eeqa
and
\beqa
( m_r- m_{0r})\frac{\mu_0}{m_{\rm u}}(\ye-Y_{\rm e0}) &\sim& 10^{34\text{--}36}\,{\rm erg\,cm^{-2}\,s^{-1}}, \\
m_{0r} \frac{\mu-\mu_0}{m_{\rm u}}(\ye-Y_{\rm e0}) &\sim& 10^{33\text{--}34}\,{\rm erg\,cm^{-2}\,s^{-1}}.
\eeqa
Combining all these contributions, 
one obtains 
\beq
F_{{\rm dis},r} \sim 10^{36-38}\,{\rm erg\,cm^{-2}\,s^{-1}}.
\eeq
Recalling that $4\pi r^2 \sim 10^{14}\,{\rm cm^2}$, one can estimate the acoustic power 
as 
\beq
\dot{E}_{\rm aco} \sim 10^{50\text{--}52}\,{\rm erg\,s^{-1}}.
\eeq

\begin{figure}[tbp]
\centering
\includegraphics[width=\hsize]{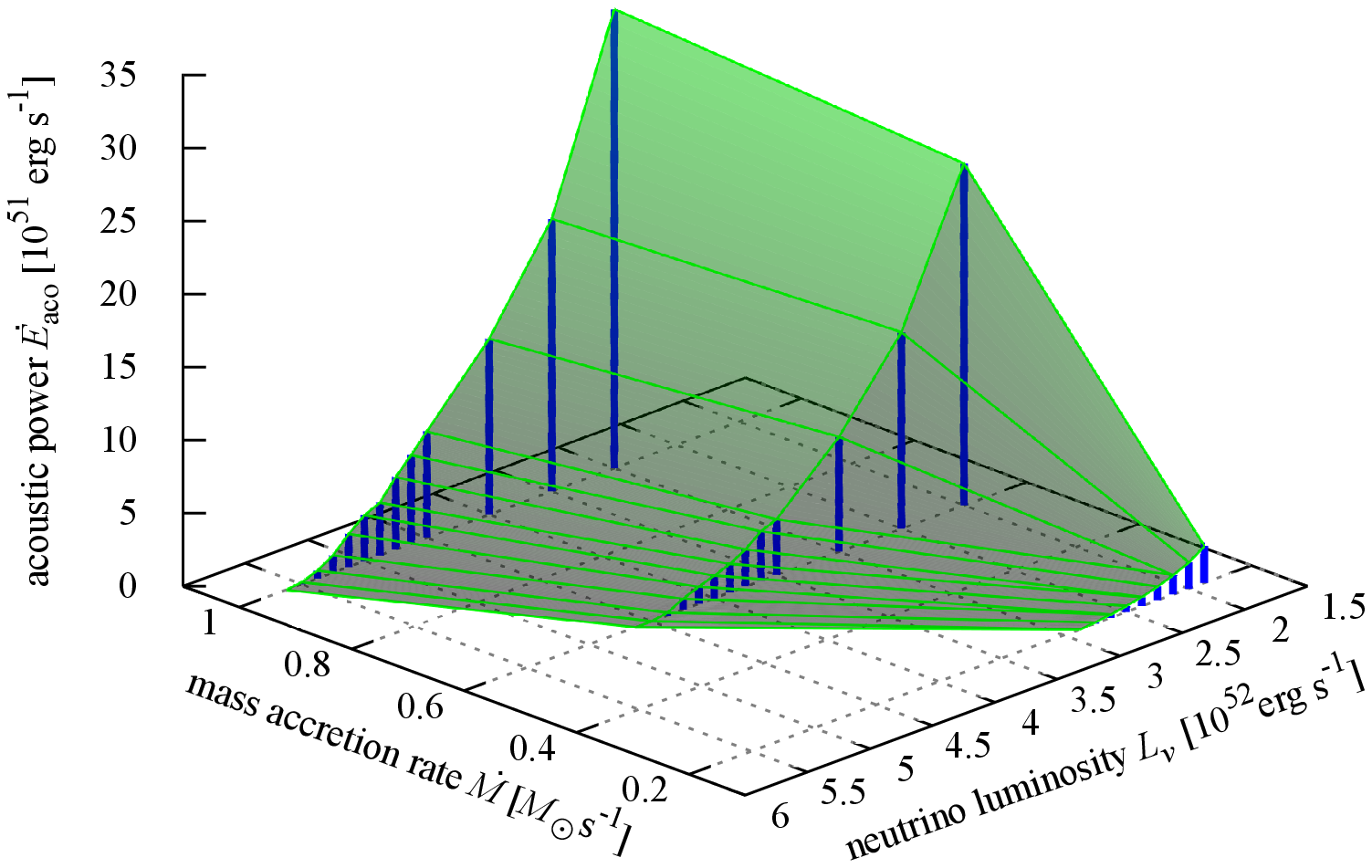}\\
\includegraphics[width=\hsize]{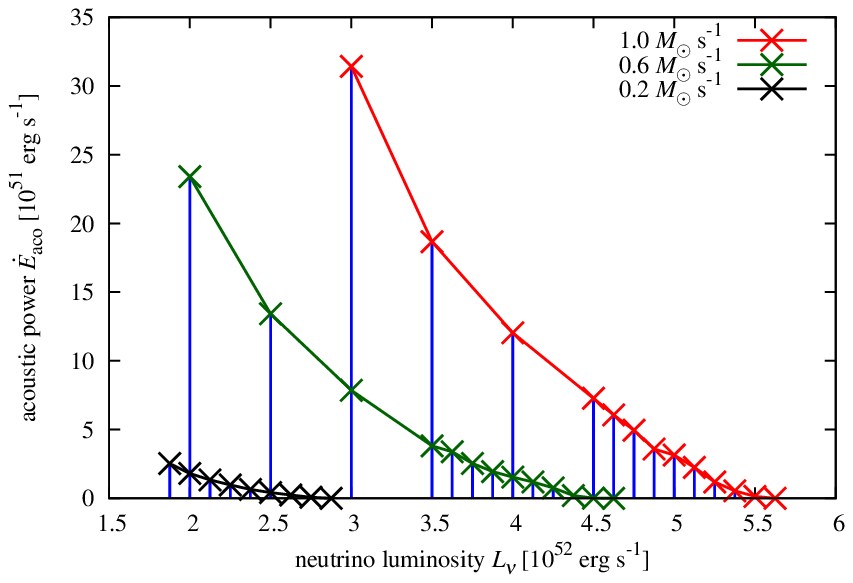}
\caption{\label{fig:1dcritene} Same as Figure \ref{fig:1dcrit} except that the vertical axes 
are the acoustic power instead of the amplitude of the acoustic wave.}
\end{figure}

Employing the acoustic power 
obtained in this way, we draw the critical surface in the parameter space spanned by $\dot{M}$, $\lnu$, and $\dot{E}_{\rm aco}$, which is 
presented in the upper panel of Figure \ref{fig:1dcritene}. 
In the lower panel, on the other hand, we also show 
three lines on the critical surface, each of which connects the results for 
the identical mass accretion rate, and which are
are projected onto the $\lnu\text{--}\dot{E}_{\rm aco}$ plane. 
It is apparent that the 
acoustic power 
required for shock revival increases as the neutrino luminosity $\lnu$ decreases. This is 
a clear indication 
that 
the acoustic power indeed contributes to shock revival.

\subsection{Energetics}

\begin{figure}[tbp]
\centering
\includegraphics[width=\hsize]{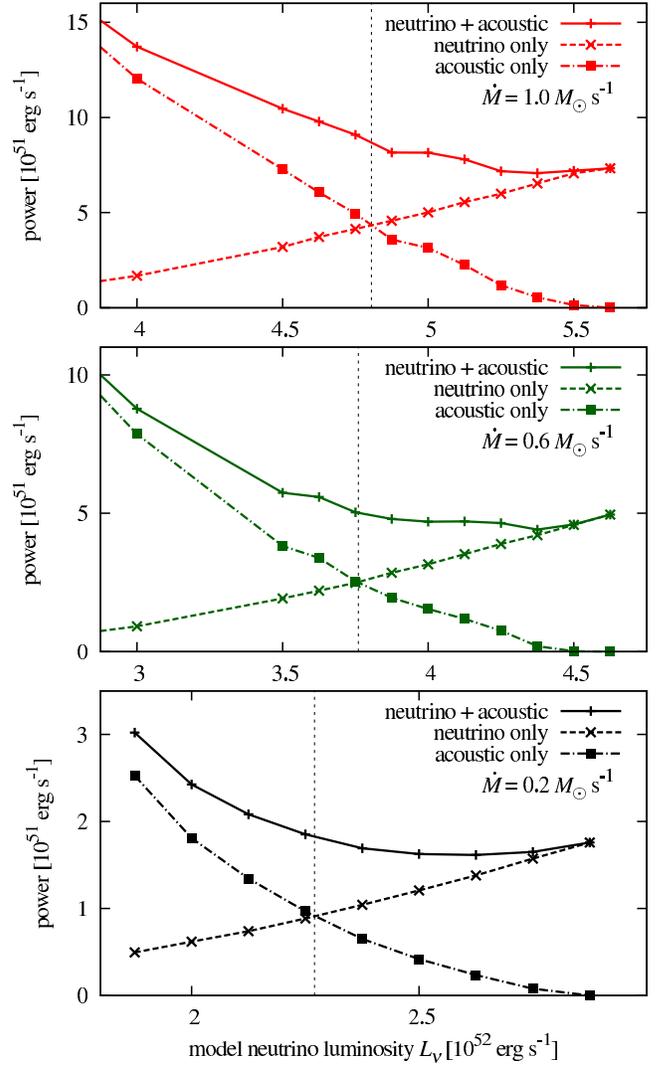}
\caption{\label{fig:energetics} Acoustic power 
(dash-dotted), the neutrino-heating rate in the gain region 
in the unperturbed state (dashed), and their sum (solid) for some models on the critical surface with different mass accretion rates. 
The upper, middle, and lower panels correspond to the mass accretion rates of 
$\mdot = 1.0\,M_\odot \,{\rm s^{-1}}$, 
$\mdot = 0.6\,M_\odot \,{\rm s^{-1}}$, and 
$\mdot = 0.2\,M_\odot \,{\rm s^{-1}}$, respectively. 
The vertical dotted lines 
indicate the points at which 
the neutrino-heating rate equals the acoustic power.}
\end{figure}

In order to 
bolster this picture, 
we investigate the energetics in more detail. 
In Figure \ref{fig:energetics} 
we compare the net neutrino-heating rate, the acoustic power, and the sum of them (or the total heating rate) for some models on the critical surface. Here 
the net neutrino-heating rate is the 
volume integral of $Q$ in equation (\ref{eq:energy}) over the gain region, which is the region where the heating by neutrino absorptions dominates the cooling by neutrino emissions. As expected intuitively, more acoustic power is needed as the neutrino luminosity decreases. 
It is found that the decrease in net neutrino-heating rate 
is almost compensated for by the increase in acoustic power, 
and the total heating rate does not change significantly for models in which the neutrino-heating rate dominates the acoustic power. 
One may be tempted to think that the explosion occurs 
if the total heating rate exceeds 
a certain threshold determined by the mass accretion rate, but this is not the case. In fact, 
for models in which 
the acoustic power 
is greater than the neutrino-heating rate, the total heating rate 
required for shock revival is no longer constant, but increases rather quickly as the neutrino luminosity decreases. 
This may imply that 
such large-power acoustic waves are inefficient in depositing their energy. 

The neutrino cooling may be responsible for the lower efficiency of the acoustic heating at large amplitudes. 
Once formed, 
the secondary shock 
waves 
raise the matter temperature, and as a consequence, enhance the neutrino cooling, since it is roughly proportional to $T^6$. As explained in appendix \ref{sec:myers}, the term $-(T-T_0)(Q/T-Q_0/T_0)$ in $D_{\rm dis}$ (see equation (\ref{eq:daco})) describes the energy loss of disturbances owing to the neutrino cooling. This effect will be more efficient for higher-power acoustic waves, since they will produce stronger secondary shock waves 
and 
lead to higher temperatures.

\begin{figure}[tbp]
\centering
\includegraphics[width=\hsize]{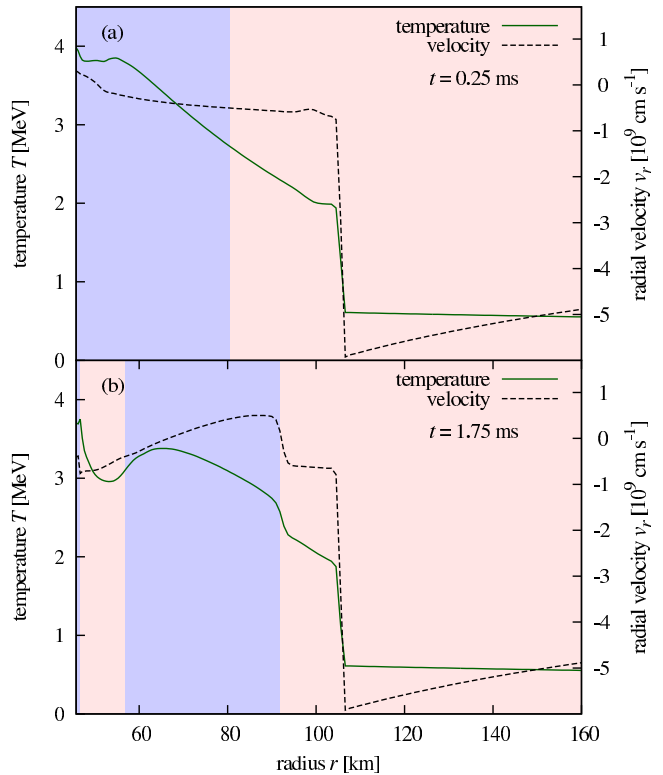}
\caption{\label{fig:secshock_temp} 
Black dashed and green solid lines 
show the radial velocity and temperature profiles, respectively, for the model on the critical surface with $\dot{M}=0.6\,M_\odot\,{\rm s^{-1}}$, $L_{\nu} = 3.0 \times 10^{52}\,{\rm erg\,s^{-1}}$, and $\delta = 0.280$ at different times. 
Red regions are the gain layers where the neutrino heating dominates cooling, whereas 
blue regions are cooling layers. 
Panels (a) and (b) track the propagation of a secondary shock, which is recognized as a discontinuous jump in the radial velocity: panel (a) is for $t=0.25\,{\rm ms}$, which is shortly after the start of the simulation; the shock is located near the inner boundary; in 
panel (b) ($t=1.75\,{\rm ms}$) 
the shock 
is shifted outward and the cooling layer trails it. 
}
\end{figure}

\begin{figure}[tbp]
\centering
\includegraphics[width=\hsize]{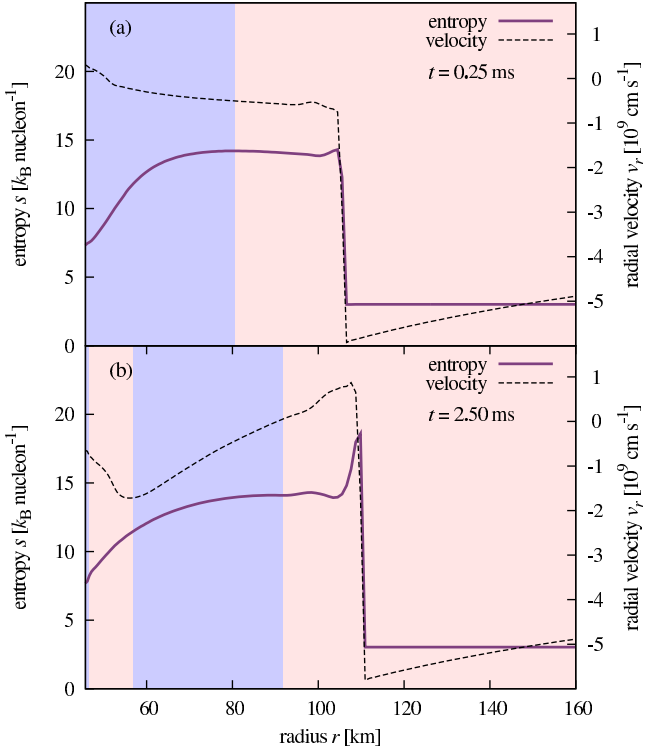}
\caption{\label{fig:mershock_ent} 
Velocity profiles (black 
dashed lines), entropy profiles (violet 
solid lines), gain layer (red regions), and coling layer (blue regions) at different times 
for the same 
model 
as in Figure \ref{fig:secshock_temp}. Panel (a) shows 
the specific entropy of the unperturbed state at the same time as in panel (a) of Figure \ref{fig:secshock_temp}. Panel (b) corresponds to the moment when the secondary shock collides with the primary shock, producing some specific entropies. 
The time is slightly later than that in panel (b) of Figure \ref{fig:secshock_temp}.
}
\end{figure}

\begin{figure}[tbp]
\centering
\includegraphics[width=\hsize]{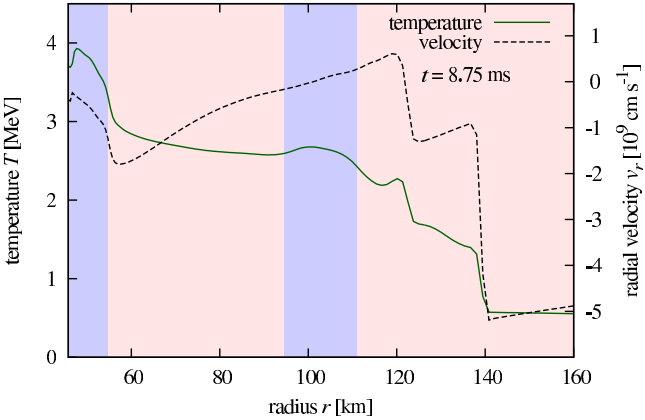}
\caption{\label{fig:distshock} 
Identical to Figure \ref{fig:secshock_temp} except that the 
time is much later, $t = 8.75\,{\rm ms}$. 
}
\end{figure}

In order to see 
this effect more quantitatively, we show the profiles of velocity, temperature, and entropy at 
different times in Figures \ref{fig:secshock_temp}, \ref{fig:mershock_ent}, and \ref{fig:distshock}. 
In these figures the cooling and heating regions are colored in blue and red, respectively. They are divided by the so-called gain radius. The cooling region sits initially closer to the inner boundary, where the temperature is higher, as shown in panels (a) of Figures \ref{fig:secshock_temp} and \ref{fig:mershock_ent}, which correspond 
to the time just after the onset of simulations. As the secondary shock propagates outward, the cooling layer 
is extended to the secondary shock at first, as seen in panel (b) of 
Figure \ref{fig:secshock_temp}. This 
is because 
the shock wave 
raises the temperature 
of the traversed matter, and as a result, enchances the neutrino cooling, thus reducing the efficiency of the acoustic heating. 
Although Burrows et al. claimed that the acoustic mechanism is 
more efficient than the neutrino-heating mechanism as 
all the energy is eventually consumed inside the primary-shock wave, this does not necessarily mean that they are used entirely for shock revival.

While cooling layers extended to just behind the secondary shock, heating layers surrounded by cooling layers sometimes appear, as shown in panels (b) of Figures \ref{fig:secshock_temp} and \ref{fig:mershock_ent} and in Figure \ref{fig:distshock}. These inner heating layers coincide with troughs in the temperature. The rarefaction of flows that follows the compression by the secondary shock leads to lower temperatures and hence suppressions of the cooling, eventually producing the heating layer there. The effect of such a inner heating layer is small, however, and the secondary shock waves enhance the cooling as a whole. We also note that the rarefaction of flows is rather sensitive to the handling of the inner boundary condition and may possibly be an artifact of such treatments.

On the other hand, not all shock heating is 
spent on the neutrino cooling, either. 
Figure \ref{fig:mershock_ent} 
shows in addition to the initial entropy distribution 
(panel (a)) 
a strong entropy production 
at the instance of the collision of the secondary shock with the primary shock 
(panel (b)). We note that this time 
is 
slightly later than the time in 
panel (b) of Figure \ref{fig:secshock_temp}. 
Since the primary shock is normally located far from the gain radius, 
the energy deposition associated with the collision 
does not 
lead to high enough 
temperatures 
for cooling to surpass heating, 
and the provided energy is not spent on neutrino emissions. 
If the primary shock is distant enough, 
the secondary shock ceases to convert the heating region into the cooling region 
well before it hits the primary shock, as demonstrated in 
Figure \ref{fig:distshock}. 
This is again simply because 
the temperature 
does not become 
high enough, and this 
is the reason why most of the acoustic power 
is still available for shock revival, even though some of the deposited energy is spent on neutrino emissions.

In addition to the neutrino cooling we just discussed, the reflection of secondary shock 
waves may 
also be the cause of the inefficiency in the acoustic heating at high amplitudes. 
According to the theory of the Riemann problem, reverse shocks or rarefaction waves are formed in general and propagate inward when the secondary shock waves collide with the primary shock. In our models, only the rarefaction waves are formed. As a consequence of these reflections, 
not all the power is provided to the primary shock. 
The reflected 
waves will hit the PNS and may be recycled, however. If this is really the case, the reflection of the secondary shock 
waves may not reduce the efficiency of energy deposition so much as we see here, since such recycling may not be properly taken into account in the Dirichlet-type inner boundary employed in our simulations.

\subsection{Diagnostics for Shock Revival}

When does shock revival occur? In the context of the neutrino-heating mechanism, several diagnostics have been proposed so far to predict it \citep{2005ApJ...620..861T, 2012ApJ...746..106P, 2017ApJ...834..183M}, although it is known that none of them is perfect. The purpose of this section is not to seek something better, but to determine whether they are useful in the current context. We examine two often-used diagnostics here: the timescale ratio, and the antesonic factor.

\begin{figure}[tbp]
\centering
\includegraphics[width=\hsize]{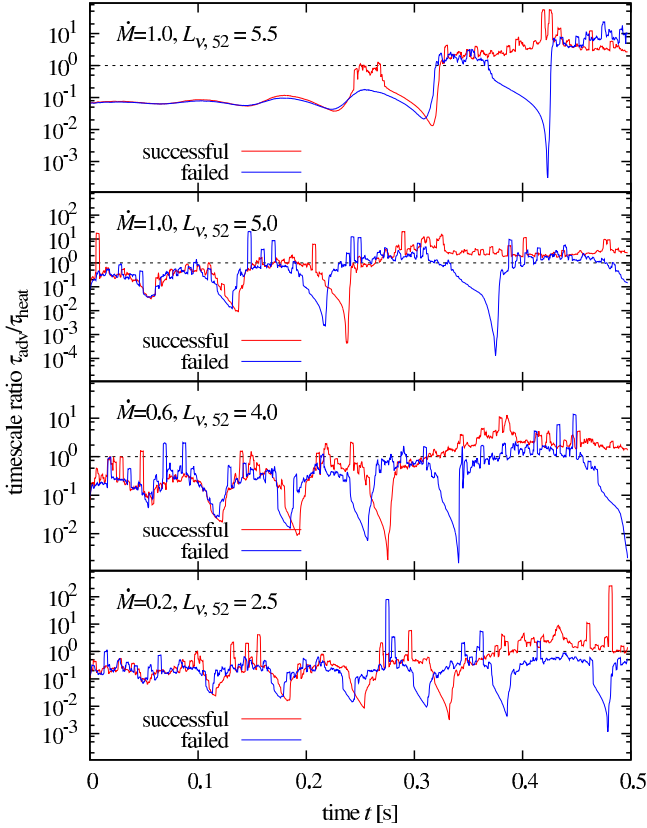}
\caption{\label{fig:runaway} 
Ratios of advection timescale to heating timescale for selected models 
slightly above and below the critical surface. 
Red lines 
show the ratios for successful models, whereas blue lines are 
for failed models. Different panels show different models. Model parameters (mass accretion rate $\dot{M}$ and neutrino luminosity $L_{\rm \nu, 52}$ in units of $M_\odot\,{\rm s^{-1}}$ and $10^{52}\,{\rm erg\,s^{-1}}$, respectively) are displayed in each 
panel. 
The timescales $\tau_{\rm adv}$ and $\tau_{\rm heat}$ are averaged over the period of $3\,{\rm ms}$. 
}
\end{figure}

One of the most frequently employed diagnostics is the ratio of advection timescale to heating timescale:
\begin{equation}
\frac{\tau_{\rm adv}}{\tau_{\rm heat}}, 
\end{equation}
where $\tau_{\rm adv} = \int_{r_{\rm gain}}^{r_{\rm shock}} {\rm d} r/|v_r|$ and $\tau_{\rm heat} = \int_{r_{\rm gain}}^{r_{\rm shock}} {\rm d} V \rho |\Phi|/Q$, with ${\rm d} V$ being 
the volume element. The radii $r_{\rm gain}$ and $r_{\rm shock}$ are the gain and shock radius, respectively. 
\citet{2005ApJ...620..861T} was the first to claim that shock revival occurs when this ratio exceeds unity, which is intuitively understandable. Although this condition was originally meant for the neutrino-heating mechanism, it may be applicable to the hybrid of acoustic power and neutrino heating studied in this paper 
if one sees it as the 
acoustic-wave-assisted neutrino-heating mechanism. 
We show in Figure \ref{fig:runaway} the timescale ratio as a function of time for both successful and failed models. 
As can be seen, the difference between the successful and unsuccessful models 
is subtle: the ratio 
exceeds one 
sometimes even for the failed models, whereas 
some times of shock oscillations occur commonly in the 
successful models before shock revivals even after the ratio reaches unity. 
We hence conclude that this diagnostic is not capable of distinguishing the successful from the failed models. 

One of the reason for this failure may be the obvious fact that only the neutrino-heating is considered in this diagnostic, although acoustic waves also contribute to the heating of matter in the present mechanism. Taking the acoustic heating into account in the discussions of heating timescales may not be so easy, however, since it is highly impulsive as the energy deposition occurs mainly when the secondary shock waves collide with the primary shock. Thus we do not pursue this issue further in this paper.

\begin{figure}[tbp]
\centering
\includegraphics[width=\hsize]{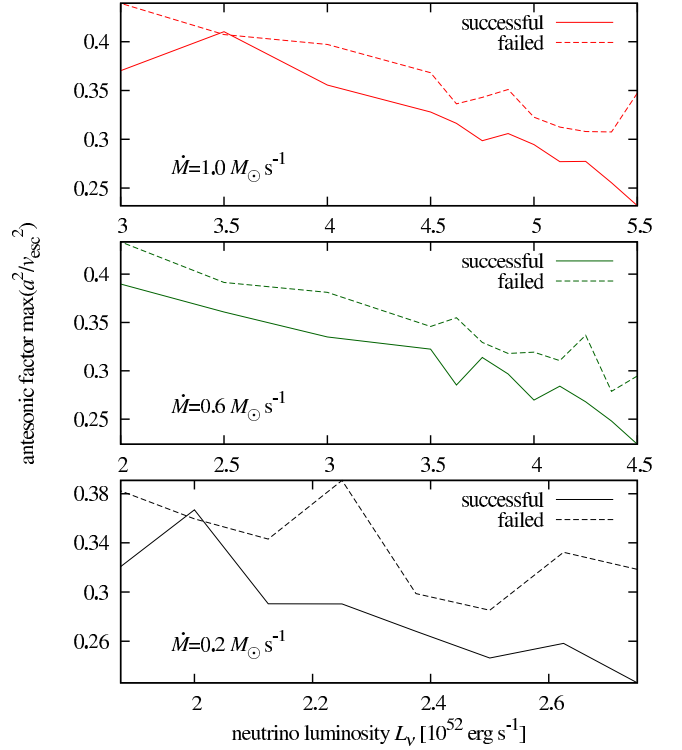}
\caption{\label{fig:antesonic} Antesonic factors ${\rm max}( a^2/v_{\rm esc}^2)$ as a function of the neutrino luminosity for some models with different mass accretion rates. Solid lines 
show the results for successful models 
slightly above the critical surface, while dashed lines 
correspond to failed models just below the critical surface. 
The mass accretion 
rates are $\dot{M}=1.0\,M_\odot\,{\rm s^{-1}}$ in the top panel, $\dot{M}=0.6\,M_\odot\,{\rm s^{-1}}$ in the middle panel, and $\dot{M}=0.2\,M_\odot\,{\rm s^{-1}}$ in the bottom panel.}
\end{figure}

According to the antesonic condition \citep{2012ApJ...746..106P}, an explosion 
should occur when the maximum value of antesonic factor, ${\rm max}( a^2/v_{\rm esc}^2)$, in the downstream of the stalled shock wave exceeds a certain 
critical value, $\sim 0.2$, where $v_{\rm esc}$ is the escape velocity. 
Figure \ref{fig:antesonic} shows 
this factor 
for some models just on 
and slightly below 
the critical surface. 
We plot the maximum value of $a^2/v_{\rm esc}^2$ 
attained by 
the time of 
shock revival in each model. 
We see from the figure that 
the maximum antesonic factor 
is larger in failed than in successful models more often than not, which is at odds with what the original theory posited. We note, however, that the antesonic condition is the condition for the steady-state solution not to exist, and as such, it may not be applicable to the current models in which the dynamical effects are also important. 
We hence conclude that the antesonic condition is not useful, either, in the mechanism considered here.

Since the diagnostics considered above 
were originally proposed in the context of the neutrino-heating mechanism, 
as mentioned repeatedly, 
it may not be surprising that 
they are not applicable to the acoustic-neutrino hybrid 
mechanism. Certainly, something better is necessary, but considering that they are known to be imperfect even in the neutrino heating, this is beyond the scope of this paper.

\section{Results in 2D}
\label{sec:2dresult}
It is true that the 1D 
models are 
convenient to understand the relevant physics, but we recall that 
the original acoustic mechanism works in 
multidimensional settings, since the $g$-mode oscillations are intrinsically non-spherical. In this section, we present 
the results of the 2D simulations 
and discuss how the dimensionality affects the critical surface.

\begin{figure*}[tbp]
\centering
\includegraphics[width=\hsize]{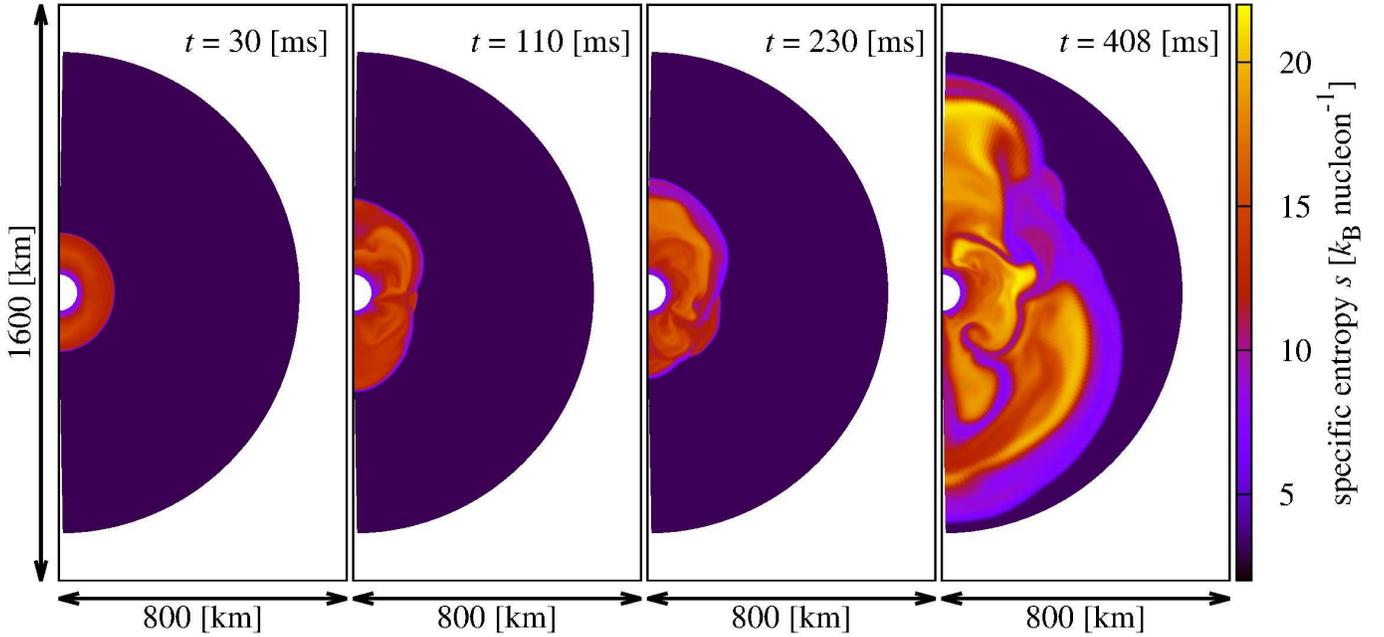}
\caption{\label{fig:entromap} Entropy 
distributions in the meridian section for the 2D model on the critical surface with $\dot{M}=0.6\,M_\odot\,{\rm s^{-1}}$, $L_\nu = 4.0 \times 10^{52}\,{\rm erg\,s^{-1}}$, and $\delta = 0.07$ 
at different times. 
The central white regions are excised from the computational domain. 
No equatorial symmetry is imposed.}
\end{figure*}

Let us 
first look at the typical 
evolution of 
2D acoustic explosion. 
In Figure \ref{fig:entromap} we show in color the entropy distributions 
in the meridian section at different times for the model with $\dot{M}=0.6\,M_\odot\,{\rm s^{-1}}$, $\lnu=4.0\times 10^{52}\,{\rm erg\,s^{-1}}$, and $\delta=0.07$, which successfully leads to shock revival. 
It is observed from the figure that the initially spherical shock is deformed preferentially along the symmetry axis by large 
plumes that are 
produced by the dipolar acoustic waves 
injected from the inner boundary in the 2D model. 
The shock morphology 
changes in time, and eventually, shock revival occurs.

\begin{figure}[tbp]
\centering
\includegraphics[width=\hsize]{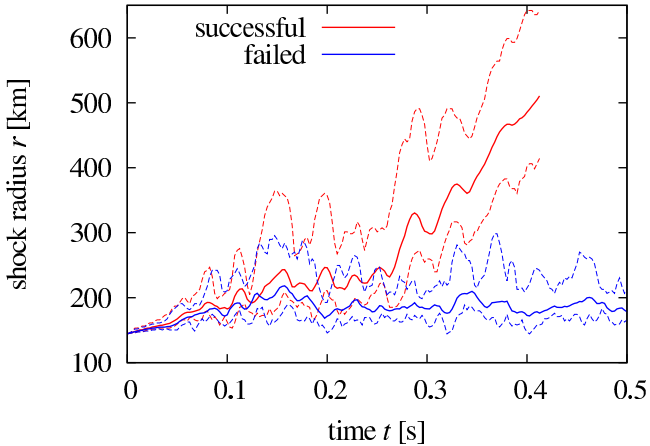}
\caption{\label{fig:2dshock} Time evolutions of the shock radii 
for the models just above and below the critical surface 
with a mass accretion rate and a neutrino luminotisy 
of $0.6\,M_\odot\,{\rm s^{-1}}$ and $4.0\times 10^{52}\,{\rm erg\,s^{-1}}$, respectively. Solid lines are the mean shock radii, 
whereas the dashed lines are the maximum and minimum 
radii. Note that the red lines 
are terminated when the maximum shock 
radius reaches the outer boundary of the computational domain.}
\end{figure}

The time evolution of the primary shock is displayed for both successful and failed models in Figure \ref{fig:2dshock}. 
Both models show 
similar evolutions in the early phases: the shock radii expand for the first $\sim 100\,{\rm ms}$ and remain almost constant for the next $\sim 100\,{\rm ms}$. The late-phase evolutions are different, on the other hand: the shock 
stays at almost the same position 
until the end of the simulation in the failed model, while 
in the successful model it rapidly expands to reach $500\,{\rm km}$ by the time of 
$\sim 410\,{\rm ms}$. We note that the difference in $\delta$ is just $0.01$ between the successful and unsuccessful models.

\begin{figure}[tbp]
\centering
\includegraphics[width=\hsize]{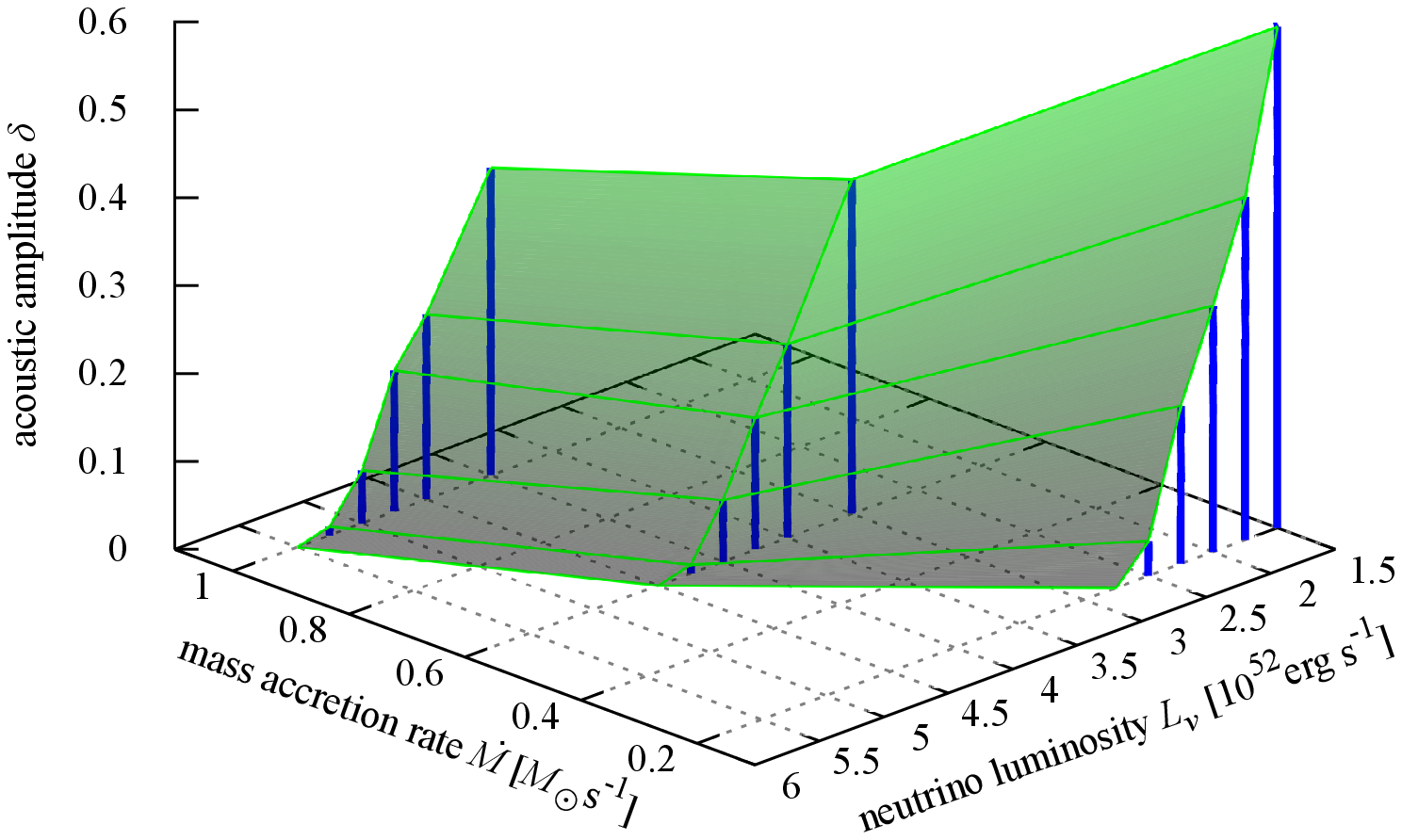}\\
\includegraphics[width=\hsize]{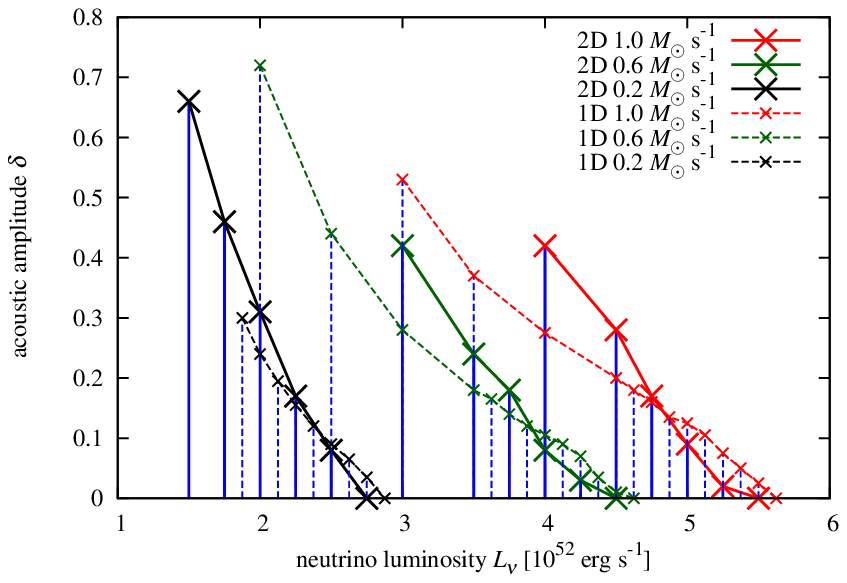}
\caption{\label{fig:2dcrit} (Upper panel) Same as Figure \ref{fig:1dcrit} 
but for the 2D 
models. (Lower panel) Lines on the critical surface for different mass accretion rates projected onto the $L_\nu$--$\delta$ plain. Line colors are the same as in Figure \ref{fig:1dcrit}. For comparison, the 1D 
counterparts are also displayed with dashed lines.}
\end{figure}

Figure \ref{fig:2dcrit} shows the 
critical surface 
in the space spanned by $\dot{M}$, $\lnu$, and $\delta$ for 2D models.
We note that 
unlike in 1D, 
the critical surface 
is not defined in a clear-cut way in 2D. This is understood from Table \ref{tab:nomono}, in which we present success or failure of shock revival for two series of models with different neutrino luminosities: $L_\nu = 4.5\times 10^{52}\,{\rm erg\,s^{-1}}$ and $4.0\times 10^{52}\,{\rm erg\,s^{-1}}$. The mass accretion rate is fixed to $\dot{M} = 1.0\,M_\odot {\rm s^{-1}}$. As $\delta$ increases in each series, a successful model appears at some point, which may be a possible critical point. For slightly higher values of $\delta$, however, we again find failure of shock revival. When we increase $\delta$ further, we eventually find only success. This is the ambiguity of the critical surface in 2D. 
The reason is that the shock revival takes place rather stochastically as a result of turbulence behind the primary shock, which is induced by convection and/or SASI. In such situations one 
may define the critical surface either as the surface below which all models fail to explode or as the surface above which all models explode. If we adopt the former, 
the critical surface passes through the points $(\dot{M},\lnu,\delta)=(1.0\,M_\odot{\rm s^{-1}},4.5\times 10^{52}\,{\rm erg\, s^{-1}},0.35)$ and $(1.0\,M_\odot{\rm s^{-1}},4.0\times 10^{52}\,{\rm erg\, s^{-1}},0.21)$, while it should pass 
through the points $(\dot{M},\lnu,\delta)=(1.0\,M_\odot{\rm s^{-1}},4.5\times 10^{52}\,{\rm erg\, s^{-1}},0.39)$ and $(1.0\,M_\odot{\rm s^{-1}},4.0\times 10^{52}\,{\rm erg\, s^{-1}},0.23)$ in the latter definition. In Figure \ref{fig:2dcrit} we adopt the former definition.

\begin{table}
\caption{Success/Failure Score-sheet\label{tab:nomono}}
\centering
\begin{tabular}{cccc}
\hline
\hline
\multicolumn{4}{c}{$\dot{M}=1.0\,M_\sun{\rm s^{-1}}$} \\
\hline
\multicolumn2c{$L_{\nu} = 4.5 \times 10^{52}\,{\rm erg\,s^{-1}}$} & \multicolumn2c{$L_{\nu} = 4.0\times 10^{52}\,{\rm erg\,s^{-1}}$} \\
\hline
$\delta = 0.40$ & successful &$\delta = 0.24$& successful \\
$\delta = 0.39$ & successful &$\delta = 0.23$& successful \\
$\delta = 0.38$ & failed         &$\delta = 0.22$& failed \\
$\delta = 0.37$ & successful &$\delta = 0.21$& successful \\
$\delta = 0.36$ & failed         &$\delta = 0.20$& failed \\
$\delta = 0.35$ & successful &$\delta = 0.19$& failed \\
$\delta = 0.34$ & failed         & & \\
$\delta = 0.33$ & failed         & & \\
\hline
\end{tabular}
\end{table}

In the lower panel of Figure \ref{fig:2dcrit} we show in the $\lnu\text{--}\delta$ plane some lines on the critical surface that have constant $\dot{M}$ for both 1D and 2D models. It is found that the critical amplitudes are not much different between 1D and 2D models, but are smaller in 2D than in 1D for large $L_\nu$ 
and vice versa 
for small $L_\nu$. 
This behavior may not be so important, however, 
since not the amplitude, but the acoustic power should be a more direct and hence a better measure for the shock revival.

\begin{figure}[tbp]
\centering
\includegraphics[width=\hsize]{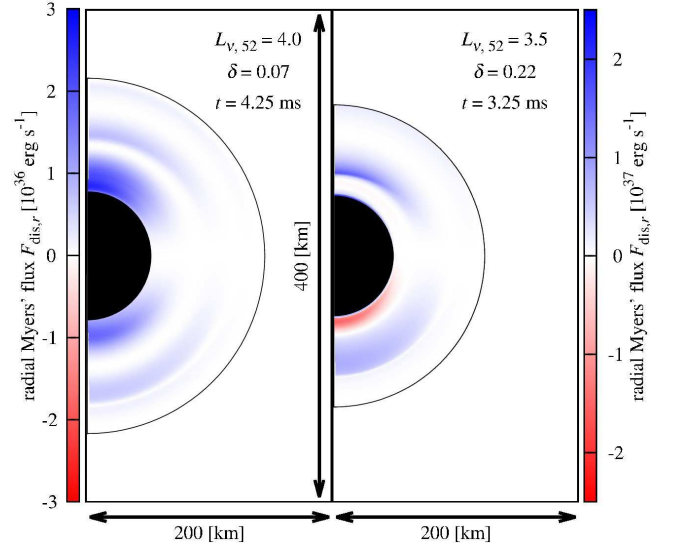}
\caption{\label{fig:2dMyerssnap} 
Radial component of the Myers flux in the meridian section for the models on the critical surface in 2D. 
Bluish colors imply that the flux is directed outward, while 
reddish ones mean that it points inward. 
The neutrino luminosities 
and the amplitudes of acoustic wave are 
$L_{\nu,\,52} = 4.0$, $\delta = 0.07$ in the left panel and 
$L_{\nu,\,52} = 3.5$, $\delta = 0.22$ in the right panel, where $L_{\nu,\,52} = L_{\nu}/(10^{52}\,{\rm erg\,s^{-1}})$. The mass accretion rate is $\dot{M}=0.6\,M_\odot\,{\rm s^{-1}}$ for both panels. 
The central black regions are excised from the computational domain. 
The outer black circles 
indicate the initial shock radii. 
The Myers flux is not shown 
outside the initial shock, 
since 
the perturbed flows are quite different from 
the unperturbed flows after shock passage.}
\end{figure}

Figure \ref{fig:2dMyerssnap} shows the radial component of the Myers flux as a contour in the meridian section. The lateral flux is negligible compared to the radial 
flux and is not shown. 
The 
black thin semi-circle 
represents the initial shock radius in each panel. 
We find that almost everwhere inside the shock, the 
Myers flux is 
directed radially outward, and as expected, it has a dipolar 
angle-dependence, i.e., it is more intense close to the symmetry axis than near the equator. 
This suggests that the Myers flux  
gives an appropriate estimate of fluxes to acoustic waves with not necessarily small amplitudes. We also note that 
the negative acoustic fluxes in the vicinity of the inner boundary mean that 
acoustic waves are reflected inward there because of the steep density gradient in the background flow.

\begin{figure}[tbp]
\centering
\includegraphics[width=\hsize]{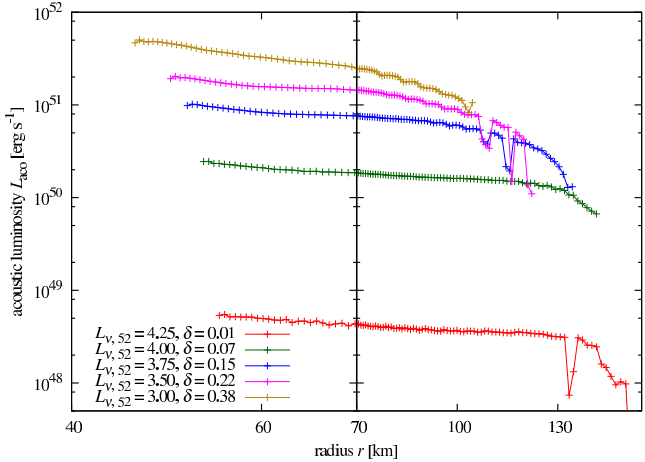}
\caption{\label{fig:2dfluxprofile} 
Acoustic luminosities $L_{\rm aco}$ 
for the 2D models with $\dot{M}=0.6\,M_\odot\,{\rm s^{-1}}$. The luminosities $L_{\rm aco}$ are 
also averaged over the period of $3\,{\rm ms}$. Line colors and legends are the same as in Figure \ref{fig:fluxprofile}.}
\end{figure}

Since the Myers fluxes are positive in almost all directions, we can employ the acoustic luminosity $L_{\rm aco}$, the surface integral of the acoustic flux, to estimate the acoustic power. 
As in 1D, we 
also take the temporal average over the oscillation period, $3\,{\rm ms}$. The resulting acoustic luminosities 
are shown in Figure \ref{fig:2dfluxprofile}. 
We note that the 
acoustic luminosity at a radius $r$ is evaluated when 
the mean radius of the acoustic wave front 
exceeds $r$. 
Since it is almost constant in radius except on the first two 
radial grid points from the inner boundary, where some adjustments are taking place, we again 
adopt its value at the third grid point as the true injected acoustic power. 

\begin{figure}[tbp]
\centering
\includegraphics[width=\hsize]{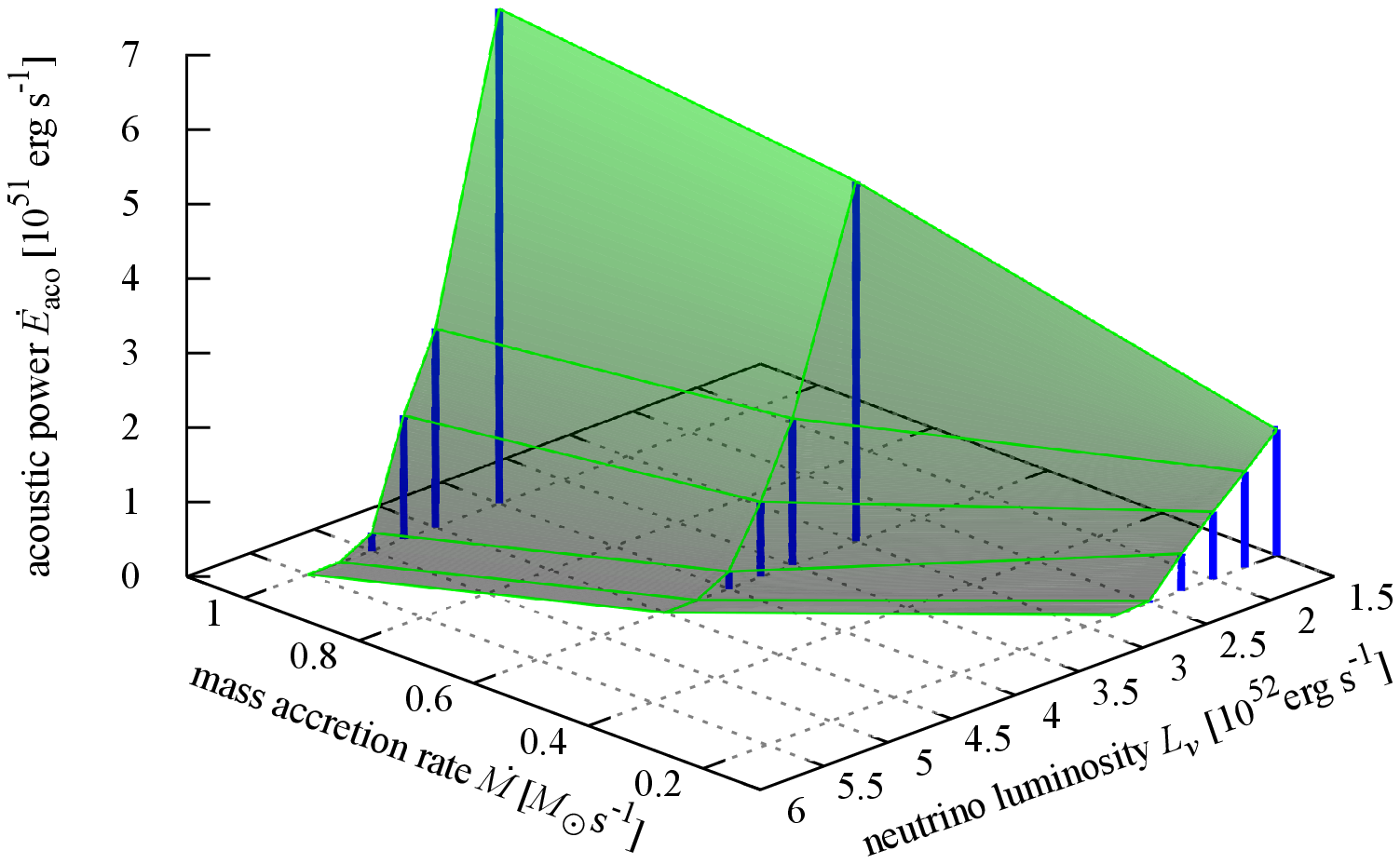}\\
\includegraphics[width=\hsize]{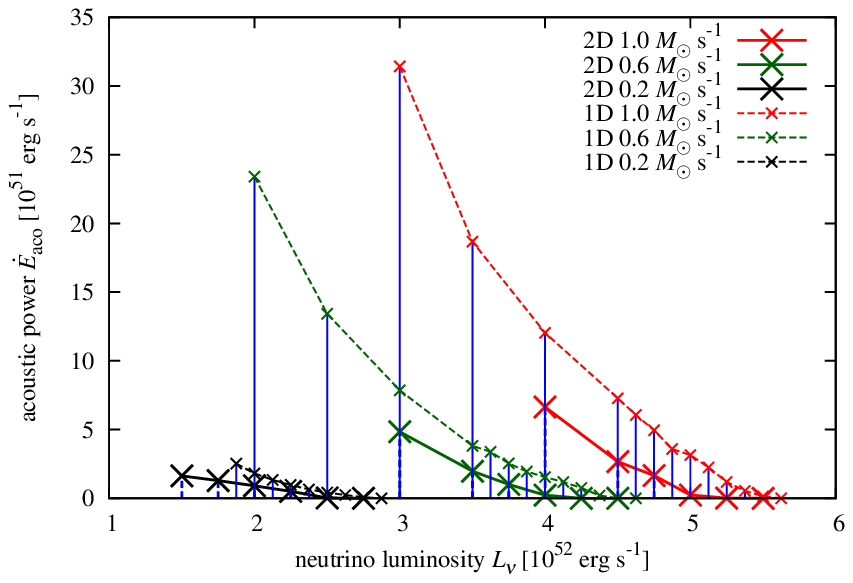}
\caption{\label{fig:2dcritene} Same as Figure \ref{fig:2dcrit} 
but for the acoustic power 
instead of the amplitude of acoustic wave.}
\end{figure}

The critical surface 
for the acoustic power 
instead of the amplitude is 
presented in Figure \ref{fig:2dcritene}. In contrast to the surface for the amplitude, 
the critical surface for the acoustic power in 2D is systematically lower than 
the 1D 
counterpart. This might to be though to be at odds with the previous findings that 
the critical surface for the acoustic amplitude is higher 
in 2D 
than 
in 1D at small $L_\nu$. The apparent contradiction 
is due to the different angular dependence of the acoustic waves between 1D and 2D. 
Since the angular integrations of the squared Legendre polynomials of our current interest are given as $\int_{-1}^1 \mathcal{P}_0(\mu)^2 \rd \mu = 2$ and $\int_{-1}^1 \mathcal{P}_1(\mu)^2 \rd \mu = 2/3$, 
the acoustic luminosities and hence powers as well are 
lower 
for the $\ell = 1$ mode than 
for the $\ell=0$ mode 
if they have the same amplitude. 

\begin{figure}[tbp]
\centering
\includegraphics[width=\hsize]{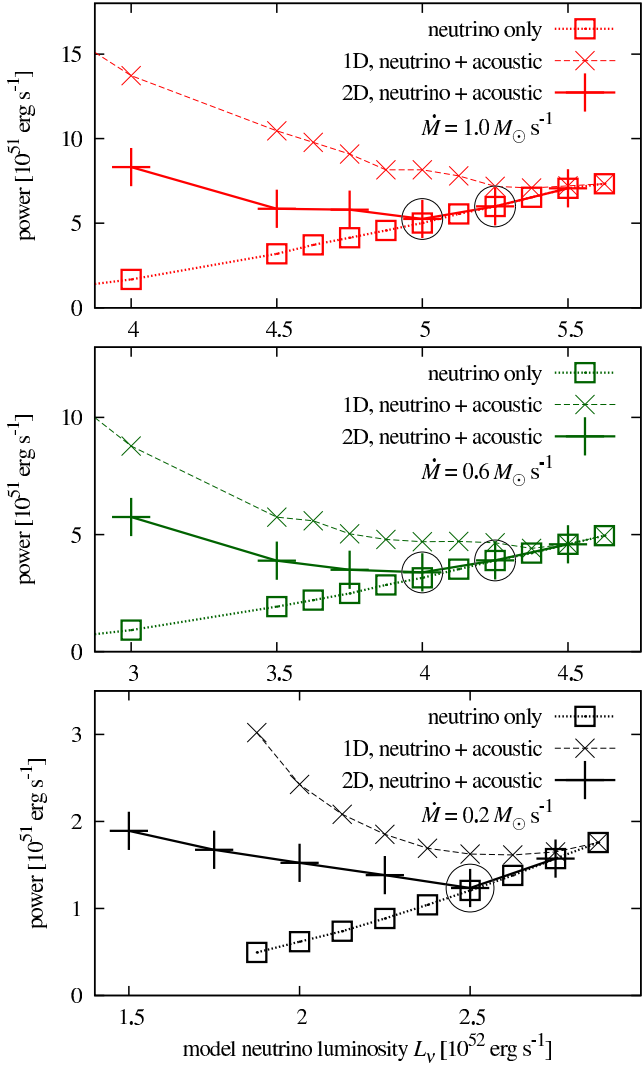}
\caption{\label{fig:1d2danalysis} Comparisons of the neutrino-heating rates (dotted lines) and 
their sums with the acoustic powers (solid lines) 
for some models on the critical surface. 
The mass accretion 
rates are 
$\dot{M}=1.0\,M_\odot\,{\rm s^{-1}}$ in the top panel, 
$\dot{M}=0.6\,M_\odot\,{\rm s^{-1}}$ in the middle panel, and 
$\dot{M}=0.2\,M_\odot\,{\rm s^{-1}}$ in the bottom panel. 
The total 
powers for the 1D counterparts with the same mass accretion rates are also shown with dashed lines. Circles indicate models whose acoustic powers are much less than neutrino-heating rates.}
\end{figure}

\begin{figure}[tbp]
\centering
\includegraphics[width=\hsize]{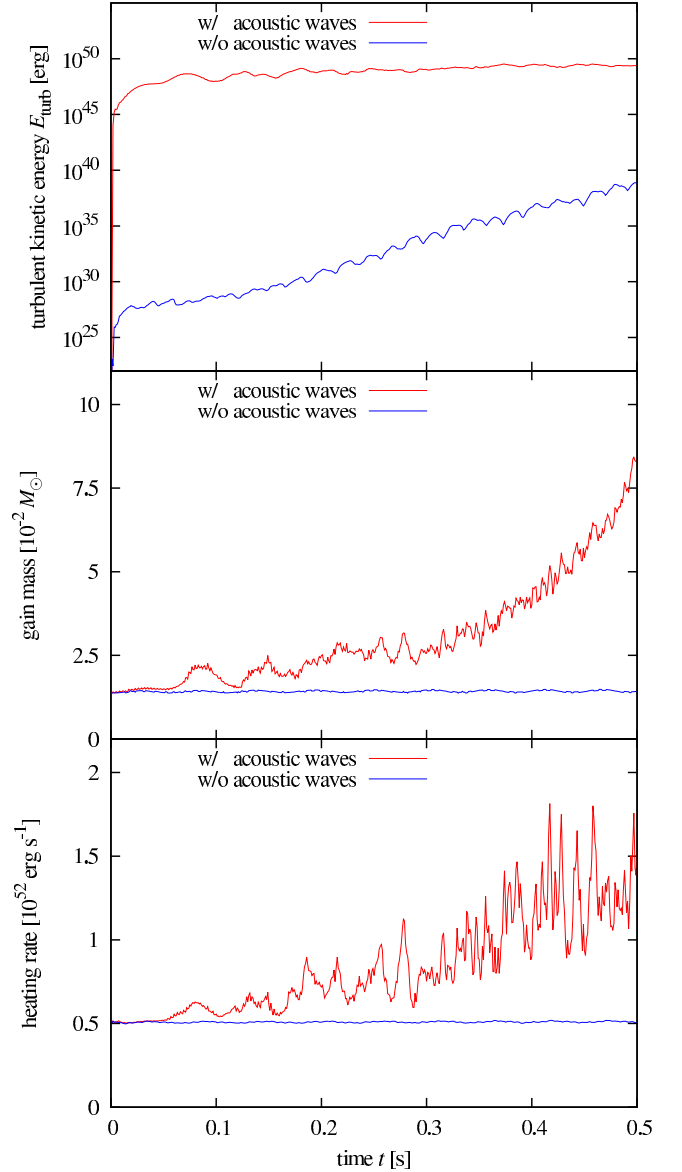}
\caption{\label{fig:enhancedheat} Comparisons of the turbulent kinetic energy (top panel), gain mass (middle panel), and the neutrino-heating rate (bottom panel) between models with (red lines) and without (blue lines) acoustic waves. For both models, $\dot{M} = 1.0\,M_\odot\,{\rm s^{-1}}$ and $\lnu = 5.0\times 10^{52}\,{\rm erg\,s^{-1}}$. Note that the model of the 
red lines lies 
on the critical surface and is one of the circeld models in Figure \ref{fig:1d2danalysis}.}
\end{figure}

We study the energetics in more detail. 
As in the previous section, we 
compare in Figure \ref{fig:1d2danalysis} the sums of the neutrino-heating rate and acoustic power with the neutrino-heating rate alone for some models on the critical surface in both 
1D and 2D. 
One can see again that shock revival in 2D requires less energy than in 1D. 
For the 
models marked with circles in the figure, 
the acoustic power needed for shock revival is 
much smaller than the neutrino-heating rate. 
In these models acoustic waves play a minor role as energy sources. Instead, 
they 
act as 
a driver of hydrodynamical instabilities, which enhance 
the neutrino heating. Since the shock revival occurs essentially by neutrino heating alone in these models, the effects of the turbulence may be estimated from the turbulent kinetic energy, dwell time, and neutrino-heating rate, which is demonstrated 
in Figure \ref{fig:enhancedheat}. The top panel of Figure \ref{fig:enhancedheat} shows the turbulent kinetic energy $E_{\rm turb}$ defined as
\beq
E_{\rm turb} = \frac{1}{2}\int_{\rm gain} \rd V \rho \left(v_{\theta}^2 + (v_r - \langle v_r \rangle)^2 \right),
\eeq
where $\langle v_r \rangle$ is the angle-averaged radial velocity, and the integral is performed over the gain region. It can be clearly seen that the injection of acoustic waves induces turbulent matter motions. 
The turbulence increases the dwell time in the gain region, and as a result, 
the gain mass, which is the mass in the gain region, also increases, as seen in the middle panel of Figure \ref{fig:enhancedheat}. This in turn raises 
the neutrino-heating rate integrated over the gain region, 
as is apparent in the lower panel of the same figure. 
Since such an enhancement is absent 
in the model without the injection of acoustic waves, we can conclude that 
acoustic waves are still playing an important role even with small amplitudes in enhancing 
the neutrino-heating rate via the fluid instability. 
Shock revival occurs essentially not via the acoustic mechanism, but via the neutrino-heating mechanism in these models.

For models with smaller $L_\nu$, 
the energy injection by acoustic waves 
plays a substantial role and it is the acoustic mechanism that gives rise to shock revival in these models. 
Here again the 
total 
powers required for shock revival are smaller in 2D than in 1D. 
The 
reason for the lower critical total powers in 2D is 
probably the enhancement of neutrino heating by the fluid instability again, although it is a minor player in this regime. In Figure \ref{fig:1d2danalysis} one also recognizes that the total critical power increases as the neutrino luminosity is decreased, which was also the case in 1D. 
Although we cannot conduct an 
analysis in 2D similar to the one given for 1D models in section \ref{sec:1dresult}, we infer that 
this is 
likely due to the 
reduced efficiency of the acoustic heating 
caused by the enhanced neutrino emissions by raised temperatures by the secondary shock waves as well as due to 
the partial reflection of acoustic waves.

\section{summary and discussions}
\label{sec:concl}
We performed 1D and 2D simulations of shock revival in the supernova core with the acoustic waves 
from 
PNS being taken into account phenomenologically. 
The 1D simulations were intended to capture the essential physics in the energy deposition by acoustic waves. For 
various combinations of mass accretion rate and neutrino luminosity, we obtained the critical 
amplitudes of an acoustic wave that devide successful from failed 
shock revival, and then 
drew the critical surface in the 
space of the neutrino luminosity $\lnu$, mass accretion rate $\dot{M}$, and acoustic amplitude $\delta$. 
In the successful models, the primary stalled shock is repeatedly 
hit by secondary shocks, into which acoustic waves steepen. 
As a consequence, the primary shock develops overstable 
oscillations and 
eventually revives. 
It is a combination of neutrino heating and acoustic power that gives rise to shock revival, however. In this sense, the mechanism considered in this paper may be referred to as ``hybrid."

In order to 
discuss the energetics quantitatively, we extended the Myers energy flux 
for finite-amplitude acoustic waves to 
incorporate neutrino contributions. 
By redrawing the critical surface with the acoustic power estimated from the extended Myers flux, we found that the sum of the neutrino heating and the acoustic power 
is almost constant on the critical surface for given mass accretion rates, with the decrease in the former being nearly compensated for by the increase in the latter. It hence appears that the critical luminosity in the neutrino-heating mechanism can be replaced by this sum in the neutrino-acoustic hybrid mechanism. For low 
neutrino luminosities, 
however, more acoustic powers 
seem to be required than to merely compensate for the decrease in the neutrino heating. This is because 
large-amplitude acoustic waves tend to become strong shock waves, resulting in higher temperatures and thus 
spending the deposited energies on 
enhanced neutrino emissions. 
Incidentally, we applied two diagnostics for shock revival, which are commonly used in the context of the neutrino-heating mechanism. 
Neither of them was found to be 
useful in the present 
mechanism.

Next, we 
ran 2D simulations. We note that the acoustic mechanism is intrinsically multidimensional, since there are no spherical $g$-mode oscillations, which are the emitters 
of acoustic waves. Although the 
critical amplitudes of acoustic wave are derived from the simulations, they are not appropriate for the comparison 
with 1D results. 
The acoustic power is more suitable, and indeed, the critical surface in 2D is always lower in the acoustic power than the 1D counterpart. 
This is due to the multidimensional fluid 
instabilities, which 
are forcibly excited by acoustic waves 
and enhance the neutrino heating. 

With the critical surface 
thus obtained, we 
revisited the 
numerical results of \citet{2006ApJ...640..878B} from the viewpoint of 
energetics. 
Although the mass accretion rate and 
neutrino luminosity 
both vary with time in \citet{2006ApJ...640..878B}, we 
detected a representative combination of them as $\dot{M} \sim 0.1\,M_\odot\,{\rm s^{-1}}$ 
and $L_\nu \sim 2.0 \times 10^{52}\,{\rm erg\,s^{-1}}$. 
The corresponding acoustic power is $\sim 4 \times 10^{51}\,{\rm erg\,s^{-1}}$, which is by inspection much larger than the critical acoustic power obtained in this paper. Our model on the critical surface 
with $\dot{M} \sim 0.2\,M_\odot\,{\rm s^{-1}}$ and $L_\nu \sim 2.0 \times 10^{52}\,{\rm erg\,s^{-1}}$ requires an acoustic power of $\sim 9\times 10^{50}\,{\rm erg\,s^{-1}}$ 
for shock revival. 
Note that the critical acoustic power decreases with the mass accretion rate. 
It is hence not surprising that Burrows et al. obtained explosions via 
the acoustic mechanism for such high acoustic powers in their simulations. 
Incidentally, the acoustic powers 
estimated by \citet{2007ApJ...665.1268Y} 
are close to the critical surface. 
On the other hand, the theory of \citet{2008MNRAS.387L..64W} predicts much smaller acoustic powers from the saturated $g$-mode oscillations, which are certainly insufficient 
for shock revival. 

The critical amplitude of density perturbation should not exceed 
unity, 
since the density 
would become negative 
otherwise. This 
may give another interesting constraint. 
Linearly extrapolating the critical surface to $\delta = 1$, 
we obtain the critical curve that runs through the points with $(\dot{M}, \lnu)=(1.0\,M_\odot\,{\rm s^{-1}},\sim 2\times 10^{52}\,{\rm erg\,s^{-1}})$, $(0.6\,M_\odot\,{\rm s^{-1}},\sim 1\times 10^{52}\,{\rm erg\,s^{-1}})$, and $(0.2\,M_\odot\,{\rm s^{-1}},\sim 1\times 10^{52}\,{\rm erg\,s^{-1}})$. 
This may imply that models 
below these luminosities do not explode even if the acoustic waves power the stalled primary shock. 
We note that we considered only sinusoidal perturbations at the inner boundary in this paper. It may hence be true that the above 
estimate may not hold for other types of perturbations with different angular modes, 
oscillation periods, and so on, but we believe that it will not be changed by the order, since the density must be positive and hence there clearly exists a maximum fluctuation amplitude, regardless of the details of disturbances. 

Although it is interesting to see that the extended Myers flux derived here serves well in estimating the energy flux of finite-amplitude acoustic waves and the critical surface obtained in the simplified settings provides useful conditions for shock revival, which seem to be consistent with realistic simulations, there are some caveats 
in the above assessment. First, our 
models neglected the turbulence in the postshock flows that should have existed 
before the injection of acoustic waves, since otherwise 
the $g$-mode oscillations could not have been 
excited in PNS in the first place. 
This problem should be also important in estimating the acoustic power, 
although it is beyond the scope of this paper. 
Second, our simulations are 2D at most. It is well known, however, that turbulence properties are 
qualitatively different between 2D and 3D. Since the inverse cascade develops in the 2D 
turbulence \citep{1967PhFl...10.1417K}, 
smaller turbulent 
eddies will be produced in 3D than in 2D \citep[e.g.,][]{2013ApJ...775...35C, 2014ApJ...786...83T, 2015ApJ...801L..24M}, 
which may result in 
reduced neutrino heating as well as 
weaker PNS 
oscillations, thus shifting the critical surface somewhat upward in 3D in terms of the acoustic power. 
Further investigations on these issues are certainly warranted. Although the neutrino heating is the most favored mechanism of CCSNe at present, we should not forget alternatives at any time.

\acknowledgments
The authors thank Masataka Ogane 
for fruitful disucussion. 
A.H. also thanks Bernhard M\"{u}ller for useful suggestion and discussion.
Numerical computations were carried out on Cray XC30 at Center for Computational Astrophysics, National Astronomical Observatory of Japan, and the supercomputer system A at High Energy Accelerator Research Organization (KEK, support by the Large Scale Simulation Program No. 14/15-17 (FY2014-2015), No. 15/16-08 (FY2015-2016), and No. 16/17-11 (FY2016-2017)). 
A.H. is supported by Advanced Leading Graduate Course for Photon Science (ALPS) in the University of Tokyo. H.N. was supported in part by JSPS Postdoctoral Fellowships for Research Abroad No. 27-348, and he was partially supported at Caltech through NSF award No. TCAN AST-1333520. This work was supported by the Grant-in-Aid for Scientific Research (B) (No. 16H03986) and Grant-in-Aid for Innovative Areas (No. 24103006) from the Ministry of Education, Culture, Sports, Science and Technology (MEXT), Japan.

\appendix

\section{Myers corollary to the theory of energy conservation} 
\label{sec:myers}
In this appendix, we extend the Myers corollary to the theory of energy conservation for finite-amplitude perturbations \citep{MYERS1986277, 1991JFM...226..383M} and derive 
equations (\ref{eq:cor})--(\ref{eq:daco}). 
Although the content 
given below is almost a summary 
of Myers' work \citep{1991JFM...226..383M}, 
except for 
the incorporation 
of neutrino heating, 
we believe that it is worth reviewing. 

Our basic equations are 
\begin{eqnarray}
\pdrv{\rho}{t} + {\bs \nabla}\ipro {\bs m} = 0 &\Leftrightarrow& C=0, \label{eq:c}\\
\pdrv{\bs v}{t} + {\bs \zeta} + {\bs \nabla}H - T{\bs \nabla}s - \frac{\mu}{m_{\rm u}}{\bs \nabla} \ye = - {\bs \nabla} \Phi + \frac{1}{\rho}{\bs M} &\Leftrightarrow& \bs L = \bs \lambda, \label{eq:m}\\
\pdrv{\rho s}{t} + {\bs \nabla}\ipro({\bs m} s) = \frac{Q - \frac{\mu}{m_{\rm u}} \gamma - {\bs v} \ipro {\bs M}}{T} &\Leftrightarrow& S = \sigma, \label{eq:s}\\
\pdrv{\rho \ye}{t} + {\bs \nabla}\ipro({\bs m}\ye) = \rho\Gamma &\Leftrightarrow& G = \gamma. \label{eq:f}
\end{eqnarray}
Equation (\ref{eq:pointmass}) is used for the gravitational potential, where $\bs m = \rho \bs v$ is the mass flux, $\bs \zeta := {\bs \omega} \bs \times {\bs v} := ({\bs \nabla} \bs \times{\bs v}) \bs\times{\bs v}$ and $\bs \omega$ is the vorticity; 
$C$, $\bs L$, $S$, and $G$ are the shorthand notations for the left-hand sides of the equations of continuity, Euler, entropy, 
and electron fractions, respectively; the corresponding right-hand side of the last three equations are denoted by $\bs \lambda$, $\sigma$, and $\gamma$, respectively. 
Equtions (\ref{eq:c})--(\ref{eq:f}) are equivalent to equations (\ref{eq:continuity})--(\ref{eq:electronfraction}) in the main body, except that 
we include the momentum transfer from the 
neutrinos, $\bs M$, in order to 
make the derivation below as general as possible. 
The relevant thermodynamic relations are
\beqa
\rd e &=& T \rd s + \frac{P}{\rho^2} \rd \rho + \frac{\mu}{m_{\rm u}} \rd \ye, \label{eq:thermene}\\
\rd h &=& T \rd s + \frac{1}{\rho} \rd P + \frac{\mu}{m_{\rm u}} \rd \ye, \label{eq:thermenth}
\eeqa
where $T$, $s$, and $h = e + P/\rho$ are the temperature, entropy, and specific enthalpy, respectively; 
$\mu$ is defined as $\mu = \mu_{\rm e}+\mu_{\rm p} - \mu_{\rm n},$ 
with $\mu_{\rm e,p,n}$ 
being the chemical potentials of electron, proton, and neutron, respectively; 
$m_{\rm u}$ is the atomic mass unit. Using these relations, we can derive from our basic 
equations (\ref{eq:c})--(\ref{eq:f}) the energy conservation law cast in the following form,
\beq
\pdrv{}{t}(\rho H - P) + {\bs \nabla}\ipro ( H {\bs m}) + {\bs m}\ipro{\bs \nabla}\Phi - Q = 0. \label{eq:H}
\eeq
Here the specific stagnation enthalpy (or the Bernoulli function) $h+\onehalf \bs v^2$ is denoted by $H$. 
We note that the identity $\bs v \ipro \bs \zeta = \bs v \ipro (\bs \omega \bs \times \bs v) = 0$ is 
used. 

Consider 
a perturbative expansion of a quantity $q$ as follows: $q({\bs r},t) = q_0({\bs r}) + \sum_{n = 1}^\infty \delta^n q_n({\bs r},t)$. The subscript $0$ denotes the 
unperturbed state with no disturbance, 
whereas the subscript $n$ 
represents the $n$th order perturbation. We note that the gravitational potential $\Phi$ is assumed to be determined by PNS in this paper 
and is hence not perturbed. 
Applying this expansion 
to each quantity in the equations given above and equating 
the terms of the same order, we obtain a sequence of equation systems that govern the perturbations at each order. 
We attach the subscript $i$ to the 
the shorthand notations introduced above, 
\beq
C_i = 0,\,\bs L_i = \bs \lambda_i,\,S_i = \sigma_i,\,\text{and}\,G_i = \gamma_i, \label{eq:shorthandorder}
\eeq
to represent the $i$th order perturbations to them. 
We also 
expand the energy-conservation equation (\ref{eq:H}):
\beqa
{\bs \nabla}\ipro (\bs m_0 H_0) + {\bs m}_0\ipro {\bs \nabla}\Phi - Q_0 &=& 0\; \text{(zero-th-order)}, \\
\pdrv{}{t}(\rho H - P)_1 + \bs \nabla \ipro (\bs m_0 H_1 + \bs m_1 H_0) + \bs m_1 \ipro \bs \nabla \Phi - Q_1 &=& 0\; \text{(first-order)}, 
\eeqa
and
\beq
\pdrv{}{t}(\rho H - P)_2 + \bs \nabla \ipro (\bs m_0 H_2 + \bs m_1 H_1 + \bs m_2 H_0) + \bs m_2 \ipro \bs \nabla \Phi - Q_2 = 0\;\text{(second-order)}.
\eeq
We rewrite equations further in the following form, expanding again 
thermodynamic quantities using the relations in equations (\ref{eq:thermene}) and (\ref{eq:thermenth}),  
and using Maxwell's relations obtained also from the same relations: 
\beqa
\left(H-Ts -\frac{\mu}{m_{\rm u}}\ye\right)_0C_0 &+& \bs m_0 \ipro (\bs L_0 - \bs \lambda_0) 
+ T_0(S_0 - \sigma_0) + \frac{\mu_0}{m_{\rm u}}(G_0 - \gamma_0) \nonumber \\
&=& 0\;\; \text{(zero-th-order)},
\eeqa
\beqa
\left(H-Ts -\frac{\mu}{m_{\rm u}}\ye\right)_0C_1 &+& \bs m_0 \ipro (\bs L_1 - \bs \lambda_1) 
 + T_0(S_1 - \sigma_1) + \frac{\mu_0}{m_{\rm u}}(G_1 - \gamma_1)  \nonumber \\
+ \left(H-Ts -\frac{\mu}{m_{\rm u}}\ye\right)_1C_0 &+& \bs m_1 \ipro (\bs L_0 - \bs \lambda_0) 
 + T_1(S_0 - \sigma_0) + \frac{\mu_1}{m_{\rm u}}(G_0 - \gamma_0) \nonumber \\
 &=& 0\;\; \text{(first-order)},
\eeqa
and
\beqa
\left(H-Ts -\frac{\mu}{m_{\rm u}}\ye\right)_0C_2 &+& \bs m_0 \ipro (\bs L_2 - \bs \lambda_2) 
+ T_0(S_2 - \sigma_2) + \frac{\mu_0}{m_{\rm u}}(G_2 - \gamma_2)  \nonumber \\
+\left(H-Ts -\frac{\mu}{m_{\rm u}}\ye\right)_2C_0 &+& \bs m_2 \ipro (\bs L_0 - \bs \lambda_0) 
 + T_2(S_0 - \sigma_0) + \frac{\mu_2}{m_{\rm u}}(G_0 - \gamma_0) \nonumber \\
&+& \pdrv{E_2}{t} + \bs \nabla \ipro \bs F_2 + D_2 = 0\;\; \text{(second-order)}, \label{eq:secondorder}
\eeqa
where $E_2$, $\bs F_2$, and $D_2$ are given as 
\beq
E_2 = \frac{P_1^2}{2\rho_0 a_0^2} + \frac{\rho_0u_1^2}{2} + \rho_1 \bs u_0 \ipro \bs u_1 + \frac{\rho_0}{2} \left\{\left(\pdrv{T}{s}\right)_{P,\ye}s_1 + \left(\pdrv{T}{\ye}\right)_{s,P}Y_{\rm e1} \right\}s_1 + \frac{\rho_0}{2 m_{\rm u}} \left\{\left(\pdrv{\mu}{\ye}\right)_{s,P} Y_{\rm e1} + \left(\pdrv{\mu}{s}\right)_{P,\ye} s_1 \right\} Y_{\rm e1}, \label{eq:seconddensity}
\eeq
\beq
\bs F_2 = (P_1 + \rho_0 \bs u_1 \ipro \bs u_0)\left(\bs u_1 + \frac{\rho_1}{\rho_0}\bs u_0\right) + \rho_0 \bs u_0 \left(s_1 T_1 + Y_{\rm e1} \frac{\mu_1}{m_{\rm u}}\right), \label{eq:secondflux}
\eeq
and
\beq
D_2 = \bs m_1 \ipro \left(\bs \zeta_1 + s_1\bs \nabla T_0 + {\ye}_1 \bs \nabla \frac{\mu_0}{m_{\rm u}}\right) 
- s_1 \bs m_0 \ipro \bs \nabla T_1 - {\ye}_1 \bs m_0 \ipro \bs \nabla \frac{\mu_1}{m_{\rm u}} 
- T_1 \sigma_1 - \frac{\mu_1}{m_{\rm u}} \gamma_1 - \bs u_1 \ipro \bs M_1 - \frac{\rho_1}{\rho_0} (\bs u_0 \ipro \bs M_1 - \bs u_1 \ipro \bs M_0).
\eeq
We note that the zeroth- and first-order equations are trivially satisfied due to equations (\ref{eq:shorthandorder}). This is not true of the second-order equation, however, and 
the last line of equation (\ref{eq:secondorder}) remains, which may 
be interpreted as the energy conservation law for the first-order perturbation for the reasons given below. We note that 
$E_2$, $\bs F_2$ and $D_2$ contain only zeroth- and 
first-order quantities.

The first three terms on 
the right-hand side of equation (\ref{eq:seconddensity}) are the 
well-known representation for the acoustic energy density in the homentropic flow. If the flow is not uniform in entropy, an extra contribution is expected from the $T\rd s$ term in the thermodynamic relations. This is the origin of the fourth term in equation (\ref{eq:seconddensity}). We note that we need to consider the product of the first-order perturbations in entropy and temperature so that it should not vanish after the average over the oscillation period, and indeed, 
$(\pa T/\pa s)_{P,\ye} s_1$ and $(\pa T/\pa \ye)_{s,P} Y_{\rm e1}$ are the changes in temperature induced by the change in entropy and electron fraction, respectively. In a similar way, 
we sholud take into account the changes in both $Y_{\rm e}$ 
and $\mu$ for the contribution from the $\mu /m_{\rm u} \rd \ye$ term. This is expressed as 
the fifth term in equation (\ref{eq:seconddensity}). 
The same considerations can be applied to 
$\bs F_2$. The first term on the right-hand side in equation (\ref{eq:secondflux}) is again the 
well-known representation for the acoustic energy flux in the homentropic flow, in which inhomogeneities
in the flow velocity are 
taken into account, whereas 
the second term originates from the changes in entropy, temperature, $\ye$, and $\mu$. 
The last term, $D_2$, is a residual that contains everything other than those included in $E_2$ and $\bs F_2$, and as such, it is admittedly the most difficult to interpret. Given the fact 
that $E_2$ and $\bs F_2$ can be regarded as the energy density and flux, however, one may interpret 
$D_2$ as a 
dissipation term. The neutrino cooling represented by the fourth term, $-T_1 \sigma_1$, in $D_2$ clearly works that way: if 
$T_1$ is positive (negative), then 
the neutrino emission will be enhanced (suppressed), leading 
to negative (positive) values of $\sigma_1 =\{ (Q- \mu\gamma/m_{\rm u} - \bs v \ipro \bs M)/T\}_1$; this makes 
$-T_1 \sigma_1$ always 
positive, implying that the neutrino cooling tends to reduce the 
perturbation energy. Other terms in $D_2$ may not be so easy to interpret, but we refer to $D_2$ as the dissipation term in this paper. 

What we have done so far is an ordinary perturbative expansion of the basic equations up to the second order. The essential idea of Myers now plays a role. We first 
remark that the energy conservation equations 
up to the second order have the same structure. In his theory, Myers 
surmised that this is true to all 
orders and recast the 
exact energy conservation law 
into the following form: 
\beqa
\left(H-Ts-\frac{\mu}{m_{\rm u}}\ye\right)_0 C &+& \bs m_0 \ipro (\bs L - \bs \lambda) 
+ T_0 (S-\sigma) + \frac{\mu_0}{m_{\rm u}}(G-\gamma) + \nonumber \\
\left(H-Ts-\frac{\mu}{m_{\rm u}}\ye\right) C_0 &+& \bs m \ipro (\bs L_0 - \bs \lambda_0) 
+ T (S_0 - \sigma_0) + \frac{\mu}{m_{\rm u}}(G_0 - \gamma_0) \nonumber \\
-\left(H-Ts -\frac{\mu}{m_{\rm u}}\ye\right)_0C_0 &-& \bs m_0 \ipro (\bs L_0 - \bs \lambda_0) 
- T_0(S_0 - \sigma_0) - \frac{\mu_0}{m_{\rm u}}(G_0 - \gamma_0) \nonumber \\
&+& \pdrv{E_{\rm dis}}{t}  + \bs \nabla \ipro \bs F_{\rm dis} + D_{\rm dis} = 0, \label{eq:fullorder}
\eeqa
where $E_{\rm dis}$, $\bs F_{\rm dis}$, and $D_{\rm dis}$ are given as 
\beqa
E_{\rm dis} &=& \rho \left(H-H_0 - T_0 (s-s_0) - \frac{\mu_0}{m_{\rm u}} (\ye - {\ye}_0)\right) 
- \bs m_0 \ipro (\bs u - \bs u_0) - (P-P_0),\\
\bs F_{\rm dis} &=& (\bs m - \bs m_0) \left(H-H_0 - T_0 (s-s_0) - \frac{\mu_0}{m_{\rm u}} (\ye - {\ye}_0)\right) 
+ \bs m_0 \left((T-T_0)(s-s_0) + \frac{\mu-\mu_0}{m_{\rm u}}(\ye - {\ye}_0)\right),\\
D_{\rm dis} &=&- (s-s_0) \bs m_0 \ipro \bs \nabla (T-T_0) - (\ye - {\ye}_0) \bs m_0 \ipro \bs \nabla \frac{\mu - \mu_0}{m_{\rm u}} 
+ (\bs m - \bs m_0) \ipro \left(\bs \zeta - \bs \zeta_0 + (s-s_0) \bs \nabla T_0 + (\ye - {\ye}_0) \bs \nabla \frac{\mu_0}{m_{\rm u}}\right) \nonumber \\
&-& (T-T_0)\left(\frac{Q}{T} - \frac{Q_0}{T_0}\right) + \frac{\mu \mu_0}{m_{\rm u}} \left(\frac{T}{\mu} - \frac{T_0}{\mu_0} \right) \left( \frac{\gamma}{T} - \frac{\gamma_0}{T_0} \right) 
- TT_0 \left(\frac{\bs m}{T} - \frac{\bs m_0}{T_0} \right) \ipro \left(\frac{\bs M}{\rho T} - \frac{\bs M_0}{\rho_0 T_0} \right). \label{eq:fullddis}
\eeqa
He then interpreted them as the density, flux, and dissipation of the energy for not necessarily small disturbances. We note that we modified the original expression to incorporate the neutrino heating in equation (\ref{eq:fullorder}). 
It is admittedly 
difficult to justify the interpretation unambiguously, but it may be somewhat comforting to point out (i) that in the limit of small 
perturbations, $E_{\rm dis}$, $\bs F_{\rm dis}$, and $D_{\rm dis}$ are reduced to 
$E_2$, $\bs F_2$, and $D_2$, respectively, and (ii) that the resulting 
equation for 
$E_{\rm dis}$, $\bs F_{\rm dis}$, and $D_{\rm dis}$ takes the conservative form. 
Ignoring the momentum transfer from neutrinos to matter, $\bs M$ and $\bs M_0$, 
which is well justified for the models considered in this paper, we finally obtain equations (\ref{eq:cor})--(\ref{eq:daco}). 
We note that the 
neutrino heating is accounted for by the first 
term in the second line of equation (\ref{eq:fullddis}).

\bibliographystyle{apj}
\bibliography{/Users/AkiraHarada/tex/bib/ref}

\begin{thebibliography}{}
\expandafter\ifx\csname natexlab\endcsname\relax\def\natexlab#1{#1}\fi

\bibitem[{{Bruenn} {et~al.}(2013){Bruenn}, {Mezzacappa}, {Hix}, {Lentz},
  {Bronson Messer}, {Lingerfelt}, {Blondin}, {Endeve}, {Marronetti}, \&
  {Yakunin}}]{2013ApJ...767L...6B}
{Bruenn}, S.~W., {Mezzacappa}, A., {Hix}, W.~R., {et~al.} 2013, \apjl, 767, L6

\bibitem[{{Bruenn} {et~al.}(2016){Bruenn}, {Lentz}, {Hix}, {Mezzacappa},
  {Harris}, {Messer}, {Endeve}, {Blondin}, {Chertkow}, {Lingerfelt},
  {Marronetti}, \& {Yakunin}}]{2016ApJ...818..123B}
{Bruenn}, S.~W., {Lentz}, E.~J., {Hix}, W.~R., {et~al.} 2016, \apj, 818, 123

\bibitem[{{Burrows} {et~al.}(2007{\natexlab{a}}){Burrows}, {Dessart}, {Ott}, \&
  {Livne}}]{2007PhR...442...23B}
{Burrows}, A., {Dessart}, L., {Ott}, C.~D., \& {Livne}, E. 2007{\natexlab{a}},
  \physrep, 442, 23

\bibitem[{{Burrows} \& {Goshy}(1993)}]{1993ApJ...416L..75B}
{Burrows}, A., \& {Goshy}, J. 1993, \apjl, 416, L75

\bibitem[{{Burrows} {et~al.}(2006){Burrows}, {Livne}, {Dessart}, {Ott}, \&
  {Murphy}}]{2006ApJ...640..878B}
{Burrows}, A., {Livne}, E., {Dessart}, L., {Ott}, C.~D., \& {Murphy}, J. 2006,
  \apj, 640, 878

\bibitem[{{Burrows} {et~al.}(2007{\natexlab{b}}){Burrows}, {Livne}, {Dessart},
  {Ott}, \& {Murphy}}]{2007ApJ...655..416B}
---. 2007{\natexlab{b}}, \apj, 655, 416

\bibitem[{Colella \& Woodward(1984)}]{Colella1984174}
Colella, P., \& Woodward, P.~R. 1984, Journal of Computational Physics, 54, 174

\bibitem[{{Couch}(2013)}]{2013ApJ...775...35C}
{Couch}, S.~M. 2013, \apj, 775, 35

\bibitem[{{Couch} \& {Ott}(2015)}]{2015ApJ...799....5C}
{Couch}, S.~M., \& {Ott}, C.~D. 2015, \apj, 799, 5

\bibitem[{{Dolence} {et~al.}(2015){Dolence}, {Burrows}, \&
  {Zhang}}]{2015ApJ...800...10D}
{Dolence}, J.~C., {Burrows}, A., \& {Zhang}, W. 2015, \apj, 800, 10

\bibitem[{{Fern{\'a}ndez}(2012)}]{2012ApJ...749..142F}
{Fern{\'a}ndez}, R. 2012, \apj, 749, 142

\bibitem[{{Hanke} {et~al.}(2012){Hanke}, {Marek}, {M{\"u}ller}, \&
  {Janka}}]{2012ApJ...755..138H}
{Hanke}, F., {Marek}, A., {M{\"u}ller}, B., \& {Janka}, H.-T. 2012, \apj, 755,
  138

\bibitem[{Harten {et~al.}(1983)Harten, Lax, \& van Leer}]{1983SIAMR...25...35H}
Harten, A., Lax, P.~D., \& van Leer, B. 1983, SIAM Review, 25, 35

\bibitem[{{Iwakami} {et~al.}(2014){Iwakami}, {Nagakura}, \&
  {Yamada}}]{2014ApJ...793....5I}
{Iwakami}, W., {Nagakura}, H., \& {Yamada}, S. 2014, \apj, 793, 5

\bibitem[{{Janka} {et~al.}(2016){Janka}, {Melson}, \&
  {Summa}}]{2016ARNPS..66..341J}
{Janka}, H.-T., {Melson}, T., \& {Summa}, A. 2016, Annual Review of Nuclear and
  Particle Science, 66, 341

\bibitem[{{Kraichnan}(1967)}]{1967PhFl...10.1417K}
{Kraichnan}, R.~H. 1967, Physics of Fluids, 10, 1417

\bibitem[{{Liebend{\"o}rfer} {et~al.}(2001){Liebend{\"o}rfer}, {Mezzacappa},
  {Thielemann}, {Messer}, {Hix}, \& {Bruenn}}]{2001PhRvD..63j3004L}
{Liebend{\"o}rfer}, M., {Mezzacappa}, A., {Thielemann}, F.-K., {et~al.} 2001,
  \prd, 63, 103004

\bibitem[{{Marek} \& {Janka}(2009)}]{2009ApJ...694..664M}
{Marek}, A., \& {Janka}, H.-T. 2009, \apj, 694, 664

\bibitem[{{Melson} {et~al.}(2015){Melson}, {Janka}, \&
  {Marek}}]{2015ApJ...801L..24M}
{Melson}, T., {Janka}, H.-T., \& {Marek}, A. 2015, \apjl, 801, L24

\bibitem[{{M{\"u}ller}(2015)}]{2015MNRAS.453..287M}
{M{\"u}ller}, B. 2015, \mnras, 453, 287

\bibitem[{{Murphy} \& {Dolence}(2017)}]{2017ApJ...834..183M}
{Murphy}, J.~W., \& {Dolence}, J.~C. 2017, \apj, 834, 183

\bibitem[{{Murphy} {et~al.}(2013){Murphy}, {Dolence}, \&
  {Burrows}}]{2013ApJ...771...52M}
{Murphy}, J.~W., {Dolence}, J.~C., \& {Burrows}, A. 2013, \apj, 771, 52

\bibitem[{Myers(1986)}]{MYERS1986277}
Myers, M. 1986, Journal of Sound and Vibration, 109, 277

\bibitem[{{Myers}(1991)}]{1991JFM...226..383M}
{Myers}, M.~K. 1991, Journal of Fluid Mechanics, 226, 383

\bibitem[{{Nagakura} {et~al.}(2014){Nagakura}, {Sumiyoshi}, \&
  {Yamada}}]{2014ApJS..214...16N}
{Nagakura}, H., {Sumiyoshi}, K., \& {Yamada}, S. 2014, \apjs, 214, 16

\bibitem[{{Nordhaus} {et~al.}(2010){Nordhaus}, {Burrows}, {Almgren}, \&
  {Bell}}]{2010ApJ...720..694N}
{Nordhaus}, J., {Burrows}, A., {Almgren}, A., \& {Bell}, J. 2010, \apj, 720,
  694

\bibitem[{{Ohnishi} {et~al.}(2006){Ohnishi}, {Kotake}, \&
  {Yamada}}]{2006ApJ...641.1018O}
{Ohnishi}, N., {Kotake}, K., \& {Yamada}, S. 2006, \apj, 641, 1018

\bibitem[{{Pejcha} \& {Thompson}(2012)}]{2012ApJ...746..106P}
{Pejcha}, O., \& {Thompson}, T.~A. 2012, \apj, 746, 106

\bibitem[{{Rampp} \& {Janka}(2000)}]{2000ApJ...539L..33R}
{Rampp}, M., \& {Janka}, H.-T. 2000, \apjl, 539, L33

\bibitem[{{Shen} {et~al.}(1998){Shen}, {Toki}, {Oyamatsu}, \&
  {Sumiyoshi}}]{1998NuPhA.637..435S}
{Shen}, H., {Toki}, H., {Oyamatsu}, K., \& {Sumiyoshi}, K. 1998, Nuclear
  Physics A, 637, 435

\bibitem[{{Sumiyoshi} \& {Yamada}(2012)}]{2012ApJS..199...17S}
{Sumiyoshi}, K., \& {Yamada}, S. 2012, \apjs, 199, 17

\bibitem[{{Sumiyoshi} {et~al.}(2005){Sumiyoshi}, {Yamada}, {Suzuki}, {Shen},
  {Chiba}, \& {Toki}}]{2005ApJ...629..922S}
{Sumiyoshi}, K., {Yamada}, S., {Suzuki}, H., {et~al.} 2005, \apj, 629, 922

\bibitem[{{Takiwaki} {et~al.}(2012){Takiwaki}, {Kotake}, \&
  {Suwa}}]{2012ApJ...749...98T}
{Takiwaki}, T., {Kotake}, K., \& {Suwa}, Y. 2012, \apj, 749, 98

\bibitem[{{Takiwaki} {et~al.}(2014){Takiwaki}, {Kotake}, \&
  {Suwa}}]{2014ApJ...786...83T}
---. 2014, \apj, 786, 83

\bibitem[{{Tamborra} {et~al.}(2014){Tamborra}, {Raffelt}, {Hanke}, {Janka}, \&
  {M{\"u}ller}}]{2014PhRvD..90d5032T}
{Tamborra}, I., {Raffelt}, G., {Hanke}, F., {Janka}, H.-T., \& {M{\"u}ller}, B.
  2014, \prd, 90, 045032

\bibitem[{{Thompson} {et~al.}(2003){Thompson}, {Burrows}, \&
  {Pinto}}]{2003ApJ...592..434T}
{Thompson}, T.~A., {Burrows}, A., \& {Pinto}, P.~A. 2003, \apj, 592, 434

\bibitem[{{Thompson} {et~al.}(2005){Thompson}, {Quataert}, \&
  {Burrows}}]{2005ApJ...620..861T}
{Thompson}, T.~A., {Quataert}, E., \& {Burrows}, A. 2005, \apj, 620, 861

\bibitem[{{Weinberg} \& {Quataert}(2008)}]{2008MNRAS.387L..64W}
{Weinberg}, N.~N., \& {Quataert}, E. 2008, \mnras, 387, L64

\bibitem[{{Yamasaki} \& {Yamada}(2006)}]{2006ApJ...650..291Y}
{Yamasaki}, T., \& {Yamada}, S. 2006, \apj, 650, 291

\bibitem[{{Yoshida} {et~al.}(2007){Yoshida}, {Ohnishi}, \&
  {Yamada}}]{2007ApJ...665.1268Y}
{Yoshida}, S., {Ohnishi}, N., \& {Yamada}, S. 2007, \apj, 665, 1268

\end{thebibliography}

\end{document}